\documentclass[12pt]{article}
\usepackage{amssymb}
\usepackage{amsmath}
\usepackage{eurosym}
\usepackage{comment}
\usepackage{threeparttable}
\usepackage{multirow}
\usepackage{tikz-cd}
\usepackage{neuralnetwork}
\usepackage[top=1.25in, bottom=1.4in, left=1.25in, right=1.25in]{geometry}

\usetikzlibrary{shapes,decorations,arrows,calc,arrows.meta,fit,positioning}
\tikzset{
    -Latex,auto,node distance =1 cm and 1 cm,semithick,
    state/.style ={ellipse, draw, minimum width = 0.7 cm},
    point/.style = {circle, draw, inner sep=0.04cm,fill,node contents={}},
    bidirected/.style={Latex-Latex,dashed},
    el/.style = {inner sep=2pt, align=left, sloped},
    rct/.style = {draw},
      shift left/.style ={commutative diagrams/shift left={#1}},
  shift right/.style={commutative diagrams/shift right={#1}}
}
\usepackage[utf8]{inputenc}
\usepackage{amssymb}
\usepackage{amsfonts}
\usepackage{amsmath}
\usepackage{mathptmx}
\usepackage{eurosym}
\usepackage{graphicx}
\usepackage{rotating}
\usepackage[justification=centering]{caption}
\usepackage{comment}

\usepackage{ifthen} 
\usepackage{xifthen} 

\usepackage{calc}  
\usepackage{time}  
\usepackage{clock}  

\usepackage{harvard}

\providecommand{\cite}{\citeasnoun}
\renewcommand{\cite}{\citeasnoun}

\usepackage{bibunits}

\usepackage{fancybox}
  
\usepackage{adjustbox} 

\usepackage[justification=centering]{caption} 

\usepackage{section}
\usepackage{theorem}
\usepackage{verbatim}
\providecommand{\bTentative}{}

\usepackage{eurosym} 
\usepackage{fix-cm}  
\usepackage{textcomp} 
\usepackage{pifont}  
\usepackage[T1]{fontenc}
    
\providecommand{\SpecialFonts}{
    \usepackage{calligra} 
	
    \usepackage[varumlaut]{yfonts}  
    }

\usepackage{pdfpages} 

\usepackage{gloss}
\usepackage{makeidx} 

\usepackage{enumerate}  
\usepackage{etaremune}  

\usepackage{amssymb}  
\usepackage{amsfonts}  
\usepackage{amsmath}  

\usepackage{array}
\usepackage{bbold} 
\usepackage{dsfont}  
\usepackage{fixmath} 
\usepackage{mathtools}  
\usepackage{mathrsfs} 

\usepackage{endnotes}

\usepackage{afterpage} 
\usepackage{layout} 
\usepackage{lscape} 

\usepackage{setspace} 

\usepackage{url}  

\usepackage{longtable} 

\usepackage{subfigure}

\providecommand{\paraNumbering}{\theoremstyle{change}}
\SpecialFonts
\excludecomment{Tentative}
\excludecomment{Plan}
\excludecomment{Abstract}
\excludecomment{Advanced}
\excludecomment{DP}
\excludecomment{graphicpack}
\excludecomment{ps2pdf}
\includecomment{pdflatex}
\includecomment{Alternative}
\includecomment{Old}
\includecomment{Elementary}
\includecomment{Intermediate}
\includecomment{French}
\includecomment{reversemargin}
\excludecomment{SlidesIn} \includecomment{SlidesOut} \includecomment{Slides}
\includecomment{SlidesLandscapeOut} \includecomment{SlidesLandscape}
\includecomment{AbstractF}
\includecomment{TOCLists}
\includecomment{Calculations}
\includecomment{Article}
\includecomment{Book}
\includecomment{DP-W}
\includecomment{DP-CIR}
\includecomment{DP-SSRN}
\includecomment{DP-arXiv}
\includecomment{Journal}
\includecomment{Lectures}
\includecomment{Organization}

\renewcommand{\boldsymbol}{\mathbold}  

\newcommand{\mciteA}{\citename}
\newcommand{\mciteY}{\citeyear*}

\newcommand{\cbu}{,\xspace}
\newcommand{\dbu}{.\xspace}
\newcommand{\bbibuniteconometrica}{\begin{bibunit}[econometrica] 
    \renewcommand{\cite}{\nocite*}
    \renewcommand\refname{} \renewcommand{\cbu}{} 
    \renewcommand{\dbu}{} \renewcommand{\newline}{}
    \vspace{-3.5\baselineskip}}
\newcommand{\bbibunitagsm}{\begin{bibunit}[agsm] \renewcommand{\cite}{\nocite*}
    \renewcommand\refname{} \renewcommand{\cbu}{} \renewcommand{\dbu}{} 
    \renewcommand{\newline}{}
    \vspace{-3.5\baselineskip}}
\newcommand{\ebibunit}{\putbib[refer1] \end{bibunit}}

\newcommand{\bbibunitunsrt}{\begin{bibunit}[unsrt] \renewcommand{\cite}{\nocite*}
    \renewcommand\refname{} \renewcommand{\cbu}{} \renewcommand{\dbu}{} \renewcommand{\newline}{}
    \vspace{-3.5\baselineskip}}
\newcommand{\ebibunitunsrt}{\putbib[refer1] \end{bibunit}}

\makeindex  
\makeglossary  
\makegloss

\newcommand{\NumericNumberedLists}{
\def\labelenumi{\arabic{enumi}.}
\def\theenumi{\arabic{enumi}}
\def\labelenumii{\arabic{enumii}.}
\def\theenumii{\arabic{enumii}}
\def\p@enumii{\theenumi.}
\def\labelenumiii{\arabic{enumiii}.}
\def\theenumiii{\arabic{enumiii}}
\def\p@enumiii{\theenumi.\theenumii.}
\def\labelenumiv{\arabic{enumiv}.}
\def\theenumiv{\arabic{enumiv}}
\def\p@enumiv{\p@enumiii.\theenumiii}
}

\allowdisplaybreaks

\newlength{\totalhormargin}
\newlength{\totalvermargin}

\def\todayMY{\ifcase\month\or
  January\or February\or March\or April\or May\or June\or
  July\or August\or September\or October\or November\or
  December\fi\ \number\year}

\providecommand{\varAuthors}{Jean-Marie Dufour \thanks{\ \ \DufourAddress} \\
  McGill University}

\providecommand{\standardthanksEndong}{}
\renewcommand{\standardthanksEndong}{We are grateful to Luis Calderon, Santiago Camara, Saraswata Chaudhuri, Russell Davidson, John W. Galbraith, S\'ilvia Gon\c{c}alves, Christian Gouri\'eroux, Joann Jasiak, M. Hashem Pesaran, Ke-Li Xu, and Victoria Zinde-Walsh for their valuable suggestions and comments.
 This work was supported by the William Dow Chair of Political Economy (McGill University), the Natural Sciences and Engineering Research Council 
 of Canada, the Social Sciences and Humanities Research Council of Canada, the Fonds de recherche sur la soci\'{e}t\'{e} et la culture (Qu\'{e}bec), and the research funds provided by the University of Mannheim. }

\providecommand{\DufourAddress}{William Dow Professor of Economics, McGill University,
  Centre interuniversitaire de recherche en analyse des
  organisations (CIRANO), and Centre interuniversitaire de recherche en
  \'{e}conomie quantitative (CIREQ). Mailing address:
  Department of Economics, McGill University, Leacock Building, Room 414,
  855 Sherbrooke Street West, Montr\'{e}al, Qu\'{e}bec H3A 2T7, Canada.
  TEL: (1) 514 398 6071; FAX: (1) 514 398 4800; e-mail: 
  \protect\url=jean-marie.dufour@mcgill.ca=\thinspace. Web page:
  \protect\url{http://www.jeanmariedufour.com} }

\newcommand{\sptha}{\hspace{-0.01em}}
\newcommand{\spthb}{\hspace{-0.01em}}

\newcommand{\sppr}{}
\newcommand{\theoremname}{Theorem}
\newcommand{\acknowledgementname}{Acknowledgement}
\newcommand{\algorithmname}{Algorithm}
\newcommand{\assumptionname}{Assumption}
\newcommand{\axiomname}{Axiom}
\newcommand{\casename}{Case}
\newcommand{\claimname}{Claim}
\newcommand{\conclusionname}{Conclusion}
\newcommand{\conditionname}{Condition}
\newcommand{\conjecturename}{Conjecture}
\newcommand{\corollaryname}{Corollary}
\newcommand{\criterionname}{Criterion}
\newcommand{\definitionname}{Definition}
\newcommand{\examplename}{Example}
\newcommand{\exercisename}{Exercise}
\newcommand{\lemmaname}{Lemma}
\newcommand{\notationname}{Notation}
\newcommand{\problemname}{Problem}
\newcommand{\proofname}{\sppr Proof}

\newcommand{\propertyname}{\sppr Property}
\newcommand{\propositionname}{Proposition}
\newcommand{\reflistname}{References}
\newcommand{\remarkname}{Remark}
\newcommand{\resultname}{Result}
\newcommand{\solutionname}{Solution}

\newcommand{\summaryname}{Summary}

\newenvironment{proof}[1][\sptha \proofname]{\par
  \normalfont
  \trivlist
  \item[\hskip\labelsep\scshape
    #1{.}]\ignorespaces}
    {\qed\endtrivlist\vspace{\baselineskip}}

\newenvironment{proofnoname}[1][\sptha \proofname]{\par
  \noindent
  \normalfont}%
  {\qed\vspace{\baselineskip}}

\newenvironment{proofflex}[1][\spthb \proofname]{\par
  \normalfont
  \trivlist
  \item[\hskip\labelsep\scshape
    #1{\ }]\ignorespaces}
    {\qed\endtrivlist\vspace{\baselineskip}}

\newenvironment{proofflexb}[1][\spthb \proofname]{\par
  \normalfont
  \trivlist
  \item[\hskip\labelsep\scshape
    #1{\ }]\ignorespaces}
    {\qed\endtrivlist\vspace{\baselineskip}}

\newenvironment{proofflexc}[1][\spthb \proofname]{\par
  \noindent
  \normalfont}
  {\qed\vspace{\baselineskip}}

\newenvironment{proofflexwc}[1][\spthb \proofname]{\par
  \normalfont
  \trivlist
  \item[\hskip\labelsep\scshape
    #1{\ }]\ignorespaces}

\newenvironment{proofflexd}[1][\spthb \proofname]{\par
  \noindent
  \normalfont}
  {\vspace{-0.5\baselineskip}}

\newenvironment{sec2subsec}{\renewcommand{\section}{\subsection}}{}
\newenvironment{reflist}{\quad \newline \noindent {\Large \bf \reflistname}
  \newline \quad \newline
  \begin{list}{}{\itemindent -.15in \leftmargin .15in \parsep 0in
                 \itemsep 0in} \item \vspace{-.35in} }{\end{list}}

\newenvironment{npar}{\noindent \bf{\thesection.\thenpar}}{}
\newenvironment{subnpar}{\noindent \bf{\thesubsection.\thesubnpar}}{}

\newcounter{paran}

\newcounter{npar}[section]
\newcounter{nparr}[section]

\newcounter{subnpar}[subsection]
\newcounter{subnparr}[subsection]

\DeclareRobustCommand{\qed}{%
  \ifmmode 
  \else \leavevmode\unskip\penalty9999 \hbox{}\nobreak\hfill
  \fi
  \quad\hbox{\qedsymbol}}
\newcommand{\openbox}{\leavevmode
  \hbox to.77778em{%
  \hfil\vrule
  \vbox to.675em{\hrule width.6em\vfil\hrule}%
  \vrule\hfil}}
\DeclareRobustCommand{\qeddirect}{%
  \ifmmode 
  \else \leavevmode\unskip\penalty9999 \hbox{}\nobreak\hfill
  \fi
  \quad\hbox{\qedsymbol}}
\providecommand{\qedsymbol}{\openbox} 

\newcommand{\bproofin}{\begin{proof}}
\newcommand{\eproofin}{\end{proof}}
\newcommand{\bproofend}{\begin{proofnoname}}
\newcommand{\eproofend}{\end{proofnoname}}
\newcommand{\bproofth}{\begin{proofth}}
\newcommand{\eproofth}{\end{proofth}}

\newcommand{\thsection}{\thesection}

\newcommand{\thsectioneq}{\thesection}

\newcommand{\thsec}{\thsection}
\newcommand{\thseceq}{\thsectioneq}

\renewcommand{\theequation}{{\rm \thseceq.\arabic{equation}}}

\makeatletter
   \long
\def\@makecaption#1#2{\vskip 0\p@
   \setbox\@tempboxa\hbox{#1 #2}}
\makeatother

\makeatletter

\providecommand{\l@theorem}{\@dottedtocline{1}{0em}{5em}}
\renewcommand{\l@theorem}{\@dottedtocline{1}{0em}{5em}}

\providecommand{\listttheoremnameb}{\sectitlesize List of Definitions, Assumptions, Propositions and Theorems}

\newcommand\listoftheorems{
 \section*{\listttheoremnameb
           \@mkboth{\MakeUppercase\listttheoremnameb}
           {\MakeUppercase\listttheoremnameb}}
           \@starttoc{lth}
           }
\newcommand{\listoftheoremscont}[1]{
 \providecommand{\listttheoremnameb}{#1}
 \renewcommand{\listttheoremnameb}{#1}
 \section*{\listttheoremnameb
           \@mkboth{\MakeUppercase\listttheoremnameb}
           {\MakeUppercase\listttheoremnameb}}
           \@starttoc{lth}
                      \addcontentsline{toc}{section}{\listttheoremnameb}
           }

\makeatother

\newcommand{\captionproofflex}[2]{}

\newcommand{\bAssumptionA}{\begin{assumption}}
\newcommand{\eAssumptionA}{\end{assumption}}

\providecommand{\paraNumbering}{}
\renewcommand{\paraNumbering}{}
\paraNumbering

\newtheorem{theorem}{\sptha \theoremname}[section]
\newtheorem{theoremseq}{\sptha \theoremname}
{\theorembodyfont{\normalfont}%
 \newtheorem{theoremSeqRom}{\sptha \theoremname}}
\newtheorem{acknowledgement}{\sptha \acknowledgementname}[section]
\newtheorem{algorithm}{\sptha \algorithmname}[section]
\newtheorem{assumption}{\sptha \assumptionname}[section]
\newtheorem{assumptionSec}{\sptha \assumptionname}[section]
\newtheorem{assumptionV}{\sptha \assumptionname}[section]
\newtheorem{assumptionLetter}{\sptha \assumptionname}

\newtheorem{axiom}{\sptha \axiomname}[section]
\newtheorem{case}{\sptha \casename}[section]
\newtheorem{claim}{\sptha \claimname}[section]
\newtheorem{conclusion}{\sptha \conclusionname}[section]
\newtheorem{condition}{\sptha \conditionname}[section]
\newtheorem{conjecture}{\sptha \conjecturename}[section]
\newtheorem{corollary}[theorem]{\sptha \corollaryname}
\newtheorem{criterion}{\sptha \criterionname}[section]
\newtheorem{definition}{\sptha \definitionname}[section]
\newtheorem{definitionSec}{\sptha \definitionname}[section]
{\theorembodyfont{\normalfont}%
    \newtheorem{example}{\sptha \examplename}[section]}
{\theorembodyfont{\normalfont}
    \newtheorem{exampleSec}{\sptha \examplename}[section]}
\newtheorem{exercise}{\sptha \exercisename}[section]
\newtheorem{lemma}[theorem]{\sptha \lemmaname}
\newtheorem{lemmaSec}{\sptha \lemmaname}[section]
\newtheorem{notation}{\sptha \notationname}[section]
\newtheorem{problem}{\sptha \problemname}[section]
{\theorembodyfont{\normalfont} 
    \newtheorem{proofth}{\sptha \proofname}[section]}
\newtheorem{property}[theorem]{\sptha \propertyname}
\newtheorem{proposition}[theorem]{\sptha \propositionname}
\newtheorem{propositionSec}{\sptha \propositionname}[section]
{\theorembodyfont{\normalfont} 
    \newtheorem{remark}{\sptha \remarkname}[section]}
    {\theorembodyfont{\normalfont}%
\newtheorem{result}{\sptha \resultname}[section]}
\newtheorem{solution}{\sptha \solutionname}[section]
\newenvironment{statement}{}{}
\newtheorem{summary}{\sptha \summaryname}[section]

\providecommand{\resetcountersSection}{\renewcommand{\thsec}{\thsection} 
  \renewcommand{\thseceq}{\thsectioneq} 
  \setcounter{theorem}{0} \setcounter{definition}{0} \setcounter{equation}{0}}

\newcommand{\captionproofemptynocontent}[2]{}

\newcommand{\captionproofin}[2]{}

\providecommand{\DufourAddress}{William Dow Professor of Economics, McGill University,
  Centre interuniversitaire de recherche en analyse des
  organisations (CIRANO), and Centre interuniversitaire de recherche en
  \'{e}conomie quantitative (CIREQ). Mailing address:
  Department of Economics, McGill University, Leacock Building, Room 414,
  855 Sherbrooke Street West, Montr\'{e}al, Qu\'{e}bec H3A 2T7, Canada.
  TEL: (1) 514 398 6071; FAX: (1) 514 398 4800; e-mail: 
  \protect\url=jean-marie.dufour@mcgill.ca=\thinspace. Web page:
  \protect\url{http://www.jeanmariedufour.com} }
  
\providecommand{\EndongAddress}{Department of Economics, University of Mannheim,  L7, 35, Room 124, Mannheim, Germany, 68161. TEL: (49) 621 181 1879; e-mail:
\protect\url=endong.wang@uni-mannheim.de=\thinspace. Web page:
  \protect\url{http://www.endongwang.com}}
  
\renewcommand{\EndongAddress}{Department of Economics, University of Mannheim,  L7, 3-5, Room 124, Mannheim, Germany, 68161. TEL: (49) 621 181 1879; e-mail:
\protect\url=endong.wang@uni-mannheim.de=\thinspace. Web page:
  \protect\url{http://www.endongwang.com}}

\providecommand{\varAuthors}{Jean-Marie Dufour \thanks{\ \ \DufourAddress} \\
 McGill University}
\renewcommand{\varAuthors}{Jean-Marie Dufour \thanks{\ \ \DufourAddress} \\
 McGill University \and Endong Wang\thanks{\ \ \EndongAddress} \\
  University of Mannheim}

\providecommand{\listttheoremnameb}{List of Definitions, Assumptions, Propositions 
     and Theorems}
\renewcommand{\listttheoremnameb}{List of Definitions, Assumptions, Propositions 
     and Theorems}
	 
\renewcommand{\thefootnote}{\alph{footnote}}

\def \beginenumerateT {\begin{enumerate}}
\def \endenumerateT {\end{enumerate}}

\def \beginenumerateTh[#1] {\begin{enumerate}[#1]}
\def \endenumerateTh {\end{enumerate}}

\begin{French}

\end{French}

\begin{graphicpack}

\end{graphicpack}

\begin{DP}

\end{DP}

\begin{DP-W}

\end{DP-W}

\begin{DP-CIR}

\end{DP-CIR}

\begin{DP-SSRN}

\end{DP-SSRN}

\begin{DP-arXiv}

\end{DP-arXiv}

\begin{Lectures} 

\end{Lectures}

\begin{Organization}

\end{Organization}

\begin{reversemargin}

\end{reversemargin}

\begin{Article} 

\end{Article}

\begin{Journal}

\end{Journal}

\input{tcilatex.sty}

\excludecomment{Tentative}
\input{tcilatex}
\renewcommand{\thefootnote}{\fnsymbol{footnote}}

\begin{document}

\title{ \vspace{-2cm} Transmission channels, dynamic mediation, and higher-order
impulse responses\thanks{\standardthanksEndong} }
\author{\varAuthors }
\date{\today }
\maketitle

\begin{abstract}
\noindent
Impulse responses measure the total effect of causal changes (or \textquotedblleft
shocks\textquotedblright) but do not show how that effect is divided among mediator variables. We use the recursion
between Sims and higher-order impulse responses (Dufour and Renault, 1998)
to decompose an impulse response into
the contributions of mediator variables at every propagation stage. A mediator's contribution is the product of an exposure margin, which measures how the shock moves the mediator, and a propagation margin, the multi-horizon projection coefficients that carry these movements over to the outcome. We show that multi-horizon Granger noncausality from a mediator to the outcome implies a zero propagation margin and that a mediator's contribution equals a controlled mediator-path effect. We provide inference for the contributions and a test of the propagation margin. Applying the decomposition to U.S. oil supply news shocks, we find that consumer-price movements in the first six months make a large contribution to the subsequent decline in industrial production, whereas the contribution of the federal funds rate is small throughout.
 
 
\vspace{1em}
\end{abstract}
 
 
\onehalfspacing
\vspace{-1cm}
 
\newpage
 
\thispagestyle{empty} \newpage
 
\renewcommand{\thefootnote}{\arabic{footnote}}\pagenumbering{arabic}
\setcounter{section}{0} \setcounter{page}{1}

\section{Introduction}
\label{Sec: Introduction}
 
\resetcountersSection
 
Since \mciteA{sims1980macroeconomics} (\mciteY{sims1980macroeconomics}),
impulse responses have been the standard tool for measuring how causal changes (or \textquotedblleft
shocks\textquotedblright)
affect the economy.\footnote{See, among others,
\mciteA{blanchard1989dynamic} (\mciteY{blanchard1989dynamic}),
\mciteA{gali1999technology} (\mciteY{gali1999technology}),
\mciteA{romer2004new} (\mciteY{romer2004new}),
\mciteA{uhlig2005effects} (\mciteY{uhlig2005effects}),
\mciteA{mertens2013dynamic} (\mciteY{mertens2013dynamic}), and
\mciteA{stock2018identification} (\mciteY{stock2018identification}).}
Impulse responses measure total effects, however, not how those effects are
divided among mediator variables. Many questions in macroeconomics are about transmission. For
instance, do oil-price shocks lower output through monetary tightening,
higher consumer prices, or other channels? The answer matters because the same impulse response of output can arise through different channels with different policy implications, and a small impulse response can mask large effects that offset each other. How much a mediator contributes to transmission depends on how much the shock moves the mediator and on how its movements carry over to output.

Existing tools answer only part of the transmission question.
\mciteA{bernanke1997systematic} (\mciteY{bernanke1997systematic}),
henceforth BGW, attribute much of the output decline after oil-price
shocks to systematic monetary tightening, based on a counterfactual that
holds the federal funds rate fixed. \mciteA{kilian2011does}
(\mciteY{kilian2011does}), henceforth KL, argue that this counterfactual requires large
and persistent departures from the policy rule. They instead shut down
only the direct policy response to the oil price, while retaining the
responses to inflation and real activity, and find a much smaller role
for monetary policy. Their counterfactual modifies the policy
rule, which is a one-step relation in the law
of motion. Transmission, however, is dynamic: a
variable with no effect on output one period ahead can still affect output
later by moving other variables at intermediate horizons.
\mciteA{dufour1998short} (\mciteY{dufour1998short}) formalize the
predictive analogue of this idea through multi-horizon Granger causality.
Because it concerns prediction, it does not attribute an impulse response
to mediator variables.
 
This paper provides such an attribution by decomposing an impulse response into
the contributions of mediator variables. The decomposition builds on the
multi-horizon projection representation of \mciteA{dufour1998short}
(\mciteY{dufour1998short}). We call its
coefficients multi-horizon projection coefficients. The leading
coefficient is the Sims impulse response, and we call the remaining
coefficients higher-order impulse responses.\footnote{When such
multi-horizon regressions are used to estimate impulse responses, they are
known as local projections, with the current values of the variables as
regressors of interest and their lags as controls; see
\mciteA{montiel2021local} (\mciteY{montiel2021local}).
\mciteA{wolf2021same} (\mciteY{wolf2021same}) show that local projections
and VARs estimate the same impulse responses in population. Their result
concerns the Sims impulse response; the decomposition also uses the
higher-order impulse responses.}
A mediator's contribution sums products of two margins over intermediate
horizons. The exposure margin is the shock-induced movement in the
mediator, and the propagation margin is the multi-horizon projection
coefficient that carries this movement over to the outcome. Our
contributions are fourfold.
 
\emph{First}, we use the recursion between Sims and higher-order impulse
responses to write the impulse response at a given horizon in terms of the shock-induced movements in the endogenous
variables up to an intermediate horizon, the propagation stage. Each
propagation stage thus gives an exact
decomposition of the same impulse response, and the choice of stage sets
the date up to which the movements in the mediators are traced.
 
\emph{Second}, we show that the contributions of all endogenous variables sum to the
impulse response at every propagation stage. If the propagation margin of
a mediator is zero, shock-induced movements in that mediator cannot carry
over to the outcome, and the channel through that mediator is closed. Multi-horizon Granger noncausality from a mediator
to the outcome (\mciteA{dufour1998short}, \mciteY{dufour1998short}) sets
the corresponding multi-horizon projection coefficients to zero and is
therefore a sufficient condition for a closed channel.
If the exposure margin is zero, the shock does not use the channel;
a channel can therefore be open yet unused.
 
\emph{Third}, we give the mediators' contributions a causal mediation
interpretation. Within the VAR, a mediator's contribution equals a
controlled mediator-path effect: the change in the outcome when a
controlled mediator path is replaced by the shock-induced mediator path,
holding the other endogenous variables fixed. This controlled effect does
not require the cross-world assumptions used for natural indirect effects.
 
\emph{Fourth}, we provide inference for the mediators' contributions and
a test of the propagation margin. Because a contribution is a product of the
exposure and propagation margins, the delta-method approximation is
degenerate when both margins are zero and can be inaccurate when both are
close to zero. The test of the propagation margin involves only restrictions that are
linear in the multi-horizon projection coefficients, so it avoids the
degeneracy of the nonlinear product and can be carried out by OLS.
 
We apply the decomposition to U.S.\ oil supply news shocks, identified with
the external instrument of \mciteA{kanzig2021macroeconomic}
(\mciteY{kanzig2021macroeconomic}) in a VAR close to those of BGW and
KL. The stage-specific decomposition shows how the contributions of the
mediators evolve as the shock propagates through the system. Consumer
prices respond persistently to the oil supply news shock and have a large
propagation margin, and their movements through month six account for
much of the subsequent decline in industrial production. At later
propagation stages, most of the impulse response is accounted for by the
other endogenous variables, whose movements by then include the effects
of the earlier rise in consumer prices. The contribution of the federal
funds rate, by contrast, is small at every horizon and propagation stage we consider, as the oil
supply news shock moves the federal funds rate too little and too
briefly.
 
\textit{Related literature.}
A closely related approach is the transmission channel analysis (TCA) of
\mciteA{wegner2025transmission} (\mciteY{wegner2025transmission}), which
also decomposes impulse responses into effects through mediator variables
and requires only the shock of interest to be identified. We differ in
three respects. First, TCA switches the shock on or off along nested
chains of variables and argues that mediation analysis is of limited use
under feedback. We instead intervene on a mediator path, so a mediator's
contribution is a controlled mediator-path effect, well defined under
feedback. Second, TCA
relies on a researcher-chosen transmission ordering as a computational
device. Our decomposition requires no transmission ordering of the mediator
variables, and its propagation margin
is zero under multi-horizon Granger noncausality. Third, TCA partitions paths from the
shock to the outcome; our decomposition instead allocates the
impulse response across the endogenous variables at each propagation
stage.

The controlled mediator-path interpretation builds on
\mciteA{robins1992identifiability} (\mciteY{robins1992identifiability}),
\mciteA{vanderweele2011controlled} (\mciteY{vanderweele2011controlled}),
and \mciteA{vanderweele2015explanation}
(\mciteY{vanderweele2015explanation}), and extends causal interpretations
of the impulse response, such as \mciteA{rambachan2021common}
(\mciteY{rambachan2021common}), to the mediators' contributions.
The multi-horizon representation and the recursion are due to
\mciteA{dufour1998short} (\mciteY{dufour1998short}), with large-sample
theory in \mciteA{dufour2006short} (\mciteY{dufour2006short}); see also
\mciteA{dufour1993relationship} (\mciteY{dufour1993relationship}) and
\mciteA{kuersteiner2010granger} (\mciteY{kuersteiner2010granger}).
 
On the empirical side, \mciteA{hamilton1983oil} (\mciteY{hamilton1983oil})
links oil-price increases to U.S.\ recessions, yet the cost share of energy
is too small to explain effects of that size
(\mciteA{kilian2008economic}, \mciteY{kilian2008economic}). The gap between
the direct cost of energy and the impulse response of output has motivated the search
for transmission channels such as monetary policy. The BGW hypothesis is
disputed by \mciteA{hamilton2004comment}
(\mciteY{hamilton2004comment}; reply in \mciteA{bernanke2004reply},
\mciteY{bernanke2004reply}) and KL. In theoretical models,
\mciteA{leduc2004quantitative} (\mciteY{leduc2004quantitative}) attribute
up to two-thirds of the output decline to systematic monetary policy,
whereas \mciteA{carlstrom2006oil} (\mciteY{carlstrom2006oil}) show that
this attribution depends on how neutral monetary policy is defined. Our decomposition
separates two questions in the debate over the monetary-policy channel:
whether an oil-price shock moves the federal funds rate (the exposure
margin) and whether movements in the federal funds rate carry over to
output (the propagation margin). The broader literature on monetary and
financial transmission mostly studies one margin at a time.
\mciteA{bernanke1995inside} (\mciteY{bernanke1995inside}),
\mciteA{jarocinski2020deconstructing} (\mciteY{jarocinski2020deconstructing}),
and \mciteA{bauer2023risk} (\mciteY{bauer2023risk}) ask how monetary
policy moves financial variables, that is, their exposure margin.
\mciteA{gilchrist2012credit} (\mciteY{gilchrist2012credit}) show that
credit spreads predict real activity, which concerns the propagation
margin. A mediator's contribution combines the two margins.

\textit{Outline.}
Section~\ref{sec:channels_dynamic_mediation} formulates the transmission question.
Section~\ref{Sec: Higher-order impulse responses} develops the decomposition using higher-order
impulse responses. Section~\ref{sec:med} gives the
controlled mediator-path interpretation. Section~\ref{sec:inference} provides
estimation and inference for the mediators' contributions. Section~\ref{sec:app}
presents the application, and Section~\ref{sec:con} concludes.

\section{Macroeconomic channels and dynamic mediation}
\label{sec:channels_dynamic_mediation}

This section states the transmission question and connects it to mediation analysis.

\subsection{Channel questions in applied macroeconomics}
\label{subsec:applied_var_channels}

Applied work studies transmission mainly in two ways. One approach estimates the impulse responses of the variables in the system to the shock. \mciteA{kanzig2021macroeconomic} (\mciteY{kanzig2021macroeconomic}), for example, identifies oil supply news shocks from oil futures price changes around OPEC announcements and reports that consumer prices and inflation expectations rise, industrial production falls, and the federal funds rate rises only gradually. Such impulse responses show how each variable in the dynamic system responds to the shock. How much of the impulse response of output is due to the movements in consumer prices or in the federal funds rate is a separate question.

A second approach changes part of the law of motion and simulates the system again; see \mciteA{sims2006does} (\mciteY{sims2006does}), \mciteA{mckay2023can} (\mciteY{mckay2023can}), and \mciteA{chen2023direct} (\mciteY{chen2023direct}) for examples. The attribution then depends on the counterfactual law of motion chosen by the researcher. In the oil-price debate, BGW and KL modify the equation for the federal funds rate in different ways and reach different conclusions about the role of monetary policy. Our decomposition keeps the estimated law of motion and allocates the impulse response across mediators at each propagation stage.

The counterfactual of BGW asks a natural question: how would output have responded to the oil-price shock had the federal funds rate stayed fixed? BGW attribute the gap between this counterfactual response and the estimated impulse response to systematic monetary policy. The gap depends on two margins: the exposure margin, which measures how much the shock moves the federal funds rate, and the propagation margin, which measures how movements in the federal funds rate carry over to output. It is small if either margin is small. Our decomposition reports the two margins separately, which links the transmission question to mediation analysis.

\subsection{Dynamic mediation}
\label{subsec:static_mediation_guide}

The static mediation model isolates the same two margins. Consider
\begin{align}
Y_t
&=
\beta_1 X_t+\beta_2 M_t+u_t,
\qquad
M_t
=
\gamma_1 X_t+v_t.
\label{eq:static_mediation}
\end{align}
Substitution gives $Y_t=(\beta_1+\beta_2\gamma_1)X_t+\beta_2v_t+u_t$. The indirect component $\beta_2\gamma_1$ is the product of the movement in the mediator, $\gamma_1$, and the coefficient $\beta_2$, which carries that movement over to the outcome; see \mciteA{baron1986moderator} (\mciteY{baron1986moderator}), \mciteA{mackinnon2012introduction} (\mciteY{mackinnon2012introduction}), and \mciteA{vanderweele2015explanation} (\mciteY{vanderweele2015explanation}). A small total effect $\beta_1+\beta_2\gamma_1$ can therefore coexist with large components of opposite sign.

The two margins correspond to different empirical questions. Event-study and external-instrument designs identify how a shock moves a candidate mediator, which is the exposure margin. How that movement carries over to the outcome at later horizons is the propagation margin. Evidence that an oil supply shock raises consumer prices therefore establishes exposure; a claim that consumer prices contribute to the decline in output also requires a nonzero propagation margin.

The dynamic problem preserves the product structure but adds timing. For example, if $Y_{t+1}=\beta_1X_t+\beta_2M_t+u_t$, $X_{t+1}=\alpha_1X_t+\eta_t$, and $M_{t+1}=\gamma_1X_t+v_t$, then
\begin{align}
Y_{t+2}
&=
(\beta_1\alpha_1+\beta_2\gamma_1)X_t+\beta_1\eta_t+\beta_2v_t+u_{t+1}.
\label{eq:dynamic_mediation_logic}
\end{align}
The term $\beta_2\gamma_1$ again combines the movement in the mediator induced by $X_t$, $\gamma_1$, with the coefficient $\beta_2$ that carries this movement over to the outcome. In a multivariate dynamic system, the mediator can move at several intermediate horizons, and the coefficient that carries each movement over to the outcome depends on the number of periods remaining until the horizon of interest.

Dynamic mediation therefore replaces one static product by a sum of products over intermediate horizons. Each product multiplies the shock-induced movement in the mediator at an intermediate horizon (the exposure margin) by the multi-horizon projection coefficient that carries this movement over to the outcome at the horizon of interest (the propagation margin). Summing these products up to a propagation stage gives the mediator's contribution at that propagation stage. As the propagation stage advances, the effects that earlier movements in the mediator have had on the other endogenous variables are attributed to those variables. The products at different intermediate horizons, and the contributions of different variables, can have opposite signs. As in the static model, a small impulse response can coexist with large contributions.

\subsection{A numerical illustration}
\label{subsec:num_illustration}

Consider the extreme case in which the impulse response is zero even
though a mediator takes part in transmission. Let $W_t=(Y_t,M_t,X_t)'$ follow a stable VAR$(6)$ with innovation covariance matrix equal to the identity. Table~\ref{tablegir} reports $\Phi_{yx,k}^{(h)}$, the multi-horizon projection coefficient on $X_{t+1-k}$ in the projection of $Y_{t+h}$. The first column is the Sims impulse response, and the other columns are higher-order impulse responses.

\begin{table}[ht]
\centering
\caption{Sims and higher-order impulse responses in a numerical VAR}
\label{tablegir}
\begin{threeparttable}
\begin{tabular}{c|rrrrrr}
\hline\hline
Horizon $h$
& $X_t$
& $X_{t-1}$
& $X_{t-2}$
& $X_{t-3}$
& $X_{t-4}$
& $X_{t-5}$
\\
\hline
1  &  0.000 & -0.080 & -0.200 & -0.100 & -0.040 &  0.010 \\
2  &  0.000 & -0.248 & -0.220 & -0.090 & -0.014 &  0.006 \\
3  &  0.000 & -0.214 & -0.074 &  0.025 &  0.009 & -0.001 \\
4  &  0.000 & -0.051 &  0.083 &  0.065 &  0.011 & -0.003 \\
5  &  0.000 &  0.068 &  0.027 & -0.002 & -0.011 &  0.002 \\
6  & -0.062 & -0.013 & -0.101 & -0.076 & -0.018 &  0.005 \\
7  & -0.124 & -0.104 & -0.128 & -0.059 & -0.008 &  0.000 \\
8  & -0.074 & -0.087 & -0.044 &  0.001 & -0.002 & -0.006 \\
9  &  0.044 & -0.012 &  0.029 &  0.027 & -0.003 & -0.004 \\
10 &  0.054 &  0.016 &  0.026 &  0.006 & -0.003 &  0.002 \\
\hline\hline
\end{tabular}
\begin{tablenotes}[flushleft]
\footnotesize
\item \emph{Note.}
The entry in column $X_{t+1-k}$, row $h$, is $\Phi_{yx,k}^{(h)}$, the coefficient
on $X_{t+1-k}$ in the projection of $Y_{t+h}$ on $W_t,\ldots,W_{t-5}$. The first column is the Sims impulse response.
The VAR$(6)$ is ordered as $W_t=(Y_t,M_t,X_t)'$ and has lag matrices
\begin{gather*}
\Phi_1=
\begin{bmatrix}
0.6 & 0.2 & 0.0\\
-0.2 & 0.7 & 0.4\\
0.2 & 0.0 & 0.6
\end{bmatrix},
\quad
\Phi_2=
\begin{bmatrix}
-0.4 & 0.36 & -0.08\\
0.1 & -0.5 & 0.0\\
0.0 & 0.1 & -0.2
\end{bmatrix},
\quad
\Phi_3=
\begin{bmatrix}
0.1 & 0.0 & -0.2\\
0.1 & -0.2 & 0.0\\
0.1 & 0.0 & 0.2
\end{bmatrix},
\\
\Phi_4=
\begin{bmatrix}
0.3 & 0.19 & -0.1\\
0.0 & 0.15 & 0.05\\
0.0 & 0.0 & 0.2
\end{bmatrix},
\quad
\Phi_5=
\begin{bmatrix}
0.0 & -0.1 & -0.04\\
0.0 & -0.02 & 0.0\\
0.0 & 0.03 & 0.08
\end{bmatrix},
\quad
\Phi_6=
\begin{bmatrix}
-0.1 & 0.03 & 0.01\\
0.0 & 0.0 & 0.0\\
-0.08 & 0.06 & 0.03
\end{bmatrix}.
\end{gather*}
\end{tablenotes}
\end{threeparttable}
\end{table}

The Sims impulse response of $Y$ to $X$ is zero at horizons one to five, while several higher-order impulse responses are sizable. Take horizon two. The recursion gives $\Phi_{yx,1}^{(2)}=\Phi_{yx,2}+(\Phi_1\Phi_1)_{yx}$. The first term is the direct-lag coefficient on $X_t$ in the equation for $Y_{t+2}$, $\Phi_{yx,2}=-0.08$. The only nonzero term in $(\Phi_1\Phi_1)_{yx}$ runs through the mediator: a unit movement in $X_t$ raises $M_{t+1}$ by $\Phi_{mx,1}=0.4$, and $M_{t+1}$ raises $Y_{t+2}$ by $\Phi_{ym,1}=0.2$ per unit, which gives $0.08$. The two terms cancel.

For a shock that raises $X_t$ by one unit and leaves $Y_t$ and $M_t$ unchanged, the contribution of $M$ at horizon two and propagation stage one is therefore $0.08$, that of $X$ is $-0.08$, and that of $Y$ is zero. The impulse response of $Y$ is zero, yet $M$ takes part in transmission: the shock moves $M$, the movement in $M$ carries over to $Y$, and the direct-lag term offsets it. The decomposition that shows this cancellation uses the higher-order impulse responses, which the next section defines.

\section{Higher-order impulse responses and impulse-response decomposition}
\label{Sec: Higher-order impulse responses}
 
This section defines the multi-horizon projection coefficients and uses their recursion to decompose the impulse response at each propagation stage and across the endogenous variables.
 
\subsection{VAR framework and higher-order impulse responses}
\label{subsec:mhlp_definition}
 
Let the vector of endogenous variables $W_t\in\mathbb R^K$ satisfy the
covariance-stationary VAR$(p)$
\begin{align}
W_t
&=
\Phi_1W_{t-1}
+\cdots+
\Phi_pW_{t-p}
+u_t,
\label{eq:var}
\end{align}
where $u_t$ is a zero-mean innovation with nonsingular covariance matrix
$\Sigma_u$. We omit deterministic terms. Let $\Phi_k=0$ for $k>p$.
Stability gives the moving-average representation
$W_t=\sum_{h\geq0}\Psi_hu_{t-h}$.
 
For any coefficient matrix, the subscript $im$ denotes its $(i,m)$ entry.
Thus $\Phi_{im,k}$ is the coefficient on $W_{m,t-k}$ in the equation for
$W_{i,t}$, and $\Phi_{im,k}^{(h)}$ is the $(i,m)$ entry of
$\Phi_k^{(h)}$, defined next.
 
For every $h\geq1$, the VAR implies the horizon-$h$ autoregressive representation
\begin{align}
W_{t+h}
&=
\Phi_1^{(h)}W_t
+
\Phi_2^{(h)}W_{t-1}
+\cdots+
\Phi_p^{(h)}W_{t-p+1}
+
u_t^{(h)},
\label{eq:mhlp_formal}
\end{align}
where $u_t^{(h)}=\sum_{\ell=0}^{h-1}\Psi_\ell u_{t+h-\ell}$ is orthogonal to the
information set generated by $W_t,\ldots,W_{t-p+1}$. We call
\eqref{eq:mhlp_formal} the multi-horizon projection representation and
$\{\Phi_k^{(h)}\}$ the multi-horizon projection coefficients. They satisfy
the recursion of \mciteA{dufour1998short} (\mciteY{dufour1998short}),
\begin{align}
\Phi_k^{(h+1)}
&=
\Phi_{k+1}^{(h)}
+
\Phi_1^{(h)}\Phi_k,
\qquad
\Phi_k^{(1)}=\Phi_k.
\label{eq:algrec_recursive}
\end{align}
Because $\Phi_k=0$ for $k>p$, the recursion gives
$\Phi_k^{(h)}=0$ for $k>p$ and all $h$. In particular,
$\Phi_1^{(h)}=\Psi_h$, so the leading coefficient is the Sims
impulse-response matrix. The remaining matrices
$\{\Phi_k^{(h)}\}_{k=2}^p$ are the \emph{higher-order impulse responses}.
\footnote{\mciteA{dufour1998short} (\mciteY{dufour1998short}) call
$\{\Phi_k^{(h)}\}_{k\ge1}$ generalized impulse response coefficients.
We use the term higher-order impulse responses to avoid confusion with the
different object of the same name in \mciteA{pesaran1998generalized}
(\mciteY{pesaran1998generalized}).}
 
The coefficient $\Phi_{im,k}^{(h)}$ measures how a movement in variable
$m$ at date $t+1-k$ carries over to the best linear $h$-step-ahead
forecast of variable $i$, holding fixed the other regressors in
\eqref{eq:mhlp_formal}. For $k=1$, it is an entry of the Sims
impulse-response matrix. For $k\geq2$, it is a higher-order impulse
response and measures how movements in the endogenous variables at earlier dates carry over to later outcomes. The higher-order impulse responses
can be large when the Sims impulse response is small, and they can have
the opposite sign. Because they relate later values of the variables to
earlier values, the multi-horizon projection coefficients, including the
higher-order impulse responses, provide the propagation margin.

The multi-horizon projection representation can be estimated by
regressing $W_{t+h}$ directly on $W_t$ and $p-1$ of its lags. In applications, usually only
the coefficient on $W_t$ is reported. The coefficients on the lags are the
higher-order impulse responses, which the decomposition also requires.
 
\begin{assumption}[Macroeconomic shocks]
\label{ass:svar}
The innovations in \eqref{eq:var} satisfy $u_t=\Theta_0\varepsilon_t$,
where $\Theta_0$ is nonsingular and the macroeconomic shocks
$\varepsilon_t$ are white noise with a diagonal covariance matrix.
\end{assumption}

Let $\nu_m$ denote the $m$th selection vector, the $m$th column of the
$K\times K$ identity matrix. For the $j$th shock, let
$\boldsymbol\theta_0=\Theta_0\nu_j$ denote its impact vector\footnote{We
omit the index $j$ from $\boldsymbol\theta_0$, $\boldsymbol\theta_h$ and
$\theta_{m,h}$ to simplify notation, because the shock is fixed throughout
the paper.} and $\boldsymbol\theta_h=\Phi_1^{(h)}\boldsymbol\theta_0$ its
impulse response at horizon $h\geq1$. We fix this shock and write
$\theta_{m,h}=\nu_m'\boldsymbol\theta_h$ for the impulse response of
variable $m$.

Under Assumption~\ref{ass:svar}, the shocks are mutually uncorrelated
and, because they are serially uncorrelated, also uncorrelated with past values of the endogenous variables, so $\boldsymbol\theta_h$ is the effect of the $j$th
shock on $W_{t+h}$. The diagonal covariance matrix leaves the scale of
each shock free; we fix it through $\boldsymbol\theta_0$, for example by
setting the impact response of one variable. Many SVAR studies assume
i.i.d.\ shocks with an identity covariance matrix, a special case of
Assumption~\ref{ass:svar} in which the unit variances are a
normalization. For the linear objects in this paper, serially and
mutually uncorrelated shocks are enough.

Identifying $\boldsymbol\theta_0$
may require further conditions, such as those on external instruments in
\mciteA{stock2018identification} (\mciteY{stock2018identification}); we
leave the identification scheme unrestricted. Identification determines
the shock-induced path $\{\boldsymbol\theta_k\}$ and hence the exposure
margin; the multi-horizon projection coefficients, and hence the
propagation margin, do not depend on it. The decomposition below does not
use Assumption~\ref{ass:svar} and holds for any impact vector.
 
\subsection{Recursive impulse-response decomposition}
\label{subsec:algebraic_components}
 
Iterating \eqref{eq:algrec_recursive} yields
\begin{align}
\Phi_k^{(h+1)}
&=
\Phi_{k+h}
+
\sum_{\ell=1}^{h}
\Phi_{h-\ell+1}\Phi_k^{(\ell)}.
\label{eq:algrec_formal}
\end{align}
Setting $k=1$ gives
\begin{align}
\Phi_1^{(h+1)}
&=
\Phi_{h+1}
+
\sum_{\ell=1}^{h}
\Phi_{h-\ell+1}\Phi_1^{(\ell)}.
\label{eq:sims_direct_mediated}
\end{align}
Equation~\eqref{eq:sims_direct_mediated} decomposes the horizon-$(h+1)$
Sims impulse response into the lag-$(h+1)$ autoregressive coefficient
$\Phi_{h+1}$, the VAR coefficient on $W_t$ in the equation for
$W_{t+h+1}$, which we call the direct-lag term, and terms that carry the
movement in $W_t$ over to $W_{t+h+1}$ through the endogenous variables at dates $t+1,\ldots,t+h$.
 
For a finite-order VAR, the direct-lag term is zero once $h+1>p$.
Beyond the lag length,
$\Phi_1^{(h+1)}
=\sum_{\ell=1}^{h}\Phi_{h-\ell+1}\Phi_1^{(\ell)}$,
so the entire Sims impulse response is generated by propagation through the endogenous variables at intermediate horizons.
 
The multi-horizon projection coefficients also have an intervention
interpretation. Replace $W_{t+1-k}$ by $W_{t+1-k}+\delta$, hold the other
regressors in \eqref{eq:mhlp_formal} and the innovations
$u_{t+1},\ldots,u_{t+h}$ fixed, and propagate the system forward from
date $t$ under the same law of motion. The induced change in $W_{t+h}$ is
$\Phi_k^{(h)}\delta$. For $k=1$, the perturbation changes the current values of the endogenous
variables; for $k\geq2$, it changes their values at an earlier date, while the later values $W_{t+2-k},\ldots,W_t$ are held
fixed. Equation~\eqref{eq:algrec_recursive} is a coefficient identity and
holds for any distribution of $u_t$; the innovation distribution matters
for identification, estimation, and inference.
 
\subsection{Decomposition at an intermediate propagation stage}
\label{subsec:general}
 
The recursion can be stopped at any propagation stage.
 
\begin{proposition}[Stage-specific impulse-response decomposition]
\label{prop:iterated_phi_identity}
Consider the VAR in \eqref{eq:var} and let $\Phi_1^{(0)}=I$. For any
integers $h$ and $r$ satisfying $0\leq r<h$,
\begin{align}
\Phi_1^{(h)}
&=
\sum_{k=0}^{r}
\Phi_{r-k+1}^{(h-r)}
\Phi_1^{(k)}.
\label{irfhn}
\end{align}
Equivalently, for any impact vector $\boldsymbol\theta_0$,
\begin{align}
\boldsymbol\theta_h
&=
\sum_{k=0}^{r}
\Phi_{r-k+1}^{(h-r)}
\boldsymbol\theta_k.
\label{eq:theta_recursion}
\end{align}
\end{proposition}
 
The proof is in Appendix~\ref{proofprop}.
 
In \eqref{eq:theta_recursion}, $h$ is the horizon of the impulse response
that is decomposed, and the propagation stage $r$
is the date $t+r$ up to which we follow the shock-induced movements in the
endogenous variables. Equation
\eqref{eq:theta_recursion} takes the movements
$\boldsymbol\theta_0,\ldots,\boldsymbol\theta_r$ as given and carries each
of them over to horizon $h$ with the multi-horizon projection
coefficients for the remaining $h-r$ periods. Written out for successive
propagation stages,
\begin{alignat*}{2}
r&=0:\qquad & \boldsymbol\theta_h&=\Phi_1^{(h)}\boldsymbol\theta_0,\\
r&=1: & \boldsymbol\theta_h&=\Phi_2^{(h-1)}\boldsymbol\theta_0+\Phi_1^{(h-1)}\boldsymbol\theta_1,\\
r&=2: & \boldsymbol\theta_h&=\Phi_3^{(h-2)}\boldsymbol\theta_0
+\Phi_2^{(h-2)}\boldsymbol\theta_1+\Phi_1^{(h-2)}\boldsymbol\theta_2,\\
&\;\;\vdots & &\;\;\vdots\\
r&=h-1:\quad & \boldsymbol\theta_h&=\Phi_h^{(1)}\boldsymbol\theta_0
+\Phi_{h-1}^{(1)}\boldsymbol\theta_1+\cdots+\Phi_1^{(1)}\boldsymbol\theta_{h-1}.
\end{alignat*}
At $r=0$, the whole impulse response is carried over from the impact
vector by the Sims impulse-response matrix $\Phi_1^{(h)}$. Each increase
in $r$ follows the movements in the endogenous variables one period further
and shortens the remaining propagation horizon $h-r$ by one period. At the
last propagation stage $r=h-1$, one period remains. Since
$\Phi_k^{(1)}=\Phi_k$ by definition in \eqref{eq:algrec_recursive}, the
coefficients are the VAR coefficients, and the decomposition is
\eqref{eq:sims_direct_mediated} with the horizon index shifted by one and
both sides post-multiplied by $\boldsymbol\theta_0$.

At an early propagation
stage, the decomposition asks how much of $\boldsymbol\theta_h$ is due to
what the shock does to each variable shortly after impact. At a late
propagation stage, it asks how much is due to the movements in each
variable shortly before horizon $h$. Each propagation stage gives
its own exact decomposition of $\boldsymbol\theta_h$, so contributions at
different propagation stages should not be added.
 
\subsection{Variable-specific contributions}
\label{subsec:vardec}
 
We next allocate each stage-specific decomposition across the endogenous
variables. Let $Y$ be the outcome of interest, with
selection vector $\nu_y$ and impulse response $\theta_{y,h}=\nu_y'\boldsymbol\theta_h$.
Inserting $I_K=\sum_{m=1}^{K}\nu_m\nu_m'$ into
\eqref{eq:theta_recursion} and premultiplying by $\nu_y'$ gives the
following definition.
 
\begin{definition}[Stage-specific contribution]
\label{def:sscc}
For $0\leq r<h$, define the contribution of variable $m$ to the impulse
response $\theta_{y,h}$ at propagation stage $r$ by
\begin{align}
\mathcal C_m(h,r)
&:=
\sum_{k=0}^{r}
\nu_y'
\Phi_{r-k+1}^{(h-r)}
\nu_m
\,
\theta_{m,k}.
\label{eq:local_channel}
\end{align}
\end{definition}
 
By construction,
\begin{align}
\theta_{y,h}
&=
\sum_{m=1}^{K}
\mathcal C_m(h,r),
\qquad
0\leq r<h.
\label{eq:channel_add_up}
\end{align}
The contributions of all endogenous variables sum to the impulse response
at every propagation stage.

When $m$ is a mediator, $\mathcal C_m(h,r)$ is a mediator's contribution. The
contribution of the variable on which the shock first acts partly
reflects its own persistence. The contribution of the outcome reflects
how much of the shock's effect has already carried over to the outcome by
date $t+r$ and then persists through its own lags.

Each summand in \eqref{eq:local_channel} is the product of an exposure
margin and a propagation margin. The exposure margin $\theta_{m,k}$ is the
shock-induced movement in variable $m$ at date $t+k$. The propagation
margin $\nu_y'\Phi_{r-k+1}^{(h-r)}\nu_m$ is the response of the outcome at
date $t+h$ to variable $m$ at date $t+k$, holding fixed the other values of the endogenous variables up to date
$t+r$. Unlike an impulse response, which is the response of a variable to
a shock, it is the response of one variable to another. KL read the coefficient on the oil price in the equation for
the federal funds rate in the same way, as the direct response of the
federal funds rate to the oil price with inflation and real activity held
fixed. The propagation margin extends this reading from one period to
$h-k$ periods and to any pair of variables. With the federal funds rate as
the mediator and output as the outcome, it measures how much output at
date $t+h$ changes when the federal funds rate is higher at date $t+k$,
given the other variables up to date $t+r$. Because the propagation margin
depends on the date of the movement as well as on $h$, a
mediator can contribute through an early, short-lived movement with a
large propagation margin or through a persistent movement with a small
one.

A contribution is nonzero only if, for some $k$, both margins are
nonzero. We call the channel through $m$ closed at $(h,r)$ if its
propagation margins are zero for $k=0,\ldots,r$, and unused if its
exposure margins are zero for $k=0,\ldots,r$. Whether a channel is closed
depends only on the law of motion, so a closed channel is closed for
every shock. Whether it is used depends on the shock. A mediator with a
large propagation margin makes a small contribution if the shock moves it
little, and another shock may use the same channel.
 
\subsection{Scope of the decomposition}
\label{subsec:scope}
 
\mciteA{bernanke1995inside} (\mciteY{bernanke1995inside}) describe
monetary transmission as a black box: the aggregate response is observed,
but the mechanism is not. Within the chosen system, the decomposition
reports, at a given horizon and propagation stage, the
shock-induced movements in the endogenous variables and the propagation
margins through which they carry over to the outcome.

The decomposition differs from both the forecast error variance
decomposition (FEVD) and the historical decomposition. An FEVD allocates
forecast error variance across shocks, while a historical
decomposition allocates a realized path across realizations of those
shocks. Equation~\eqref{eq:channel_add_up} instead holds the shock fixed
and allocates its impulse response across the endogenous variables at a given
propagation stage. A mediator's contribution can account
for a large share of the impulse response even if the shock accounts for
little of the forecast error variance or historical variation of the
outcome. KL, for example, use historical decompositions to show that
oil-price shocks contributed little to observed fluctuations. That
finding is compatible with a large contribution of a mediator to the
impulse response.

The decomposition takes the estimated VAR as given. Which variables enter the VAR and how many lags it includes are choices
made in
estimating the impulse response, and they affect the impulse response
itself. The contributions inherit these choices. Adding a variable such as
inflation expectations can reallocate contributions previously assigned to
consumer prices, and a non-diagonal linear transformation of $W_t$
reallocates contributions across the transformed variables. Like the
impulse response it decomposes, each contribution is therefore a statement
about the chosen system.

\section{A mediator's contribution as a controlled mediator-path effect}
\label{sec:med}

This section shows that, within the VAR, a mediator's contribution is the
effect of an intervention on the path of the mediator.\footnote{\mciteA{rambachan2021common}
(\mciteY{rambachan2021common}) give conditions under which common
time-series estimands, including impulse responses, identify causal
effects. We take such conditions as given for the impulse response and
ask what causal meaning a mediator's contribution has.}

\subsection{Controlled mediator-path interventions}

Fix a horizon $h$ and a propagation stage $r<h$. Consider two
paths of the endogenous variables through date $t+r$ that differ only in the
values of variable $m$ at dates $t,\ldots,t+r$. Let
$\mathbf m=(m_0,\ldots,m_r)'$ and
$\widetilde{\mathbf m}=(\widetilde m_0,\ldots,\widetilde m_r)'$
denote the two controlled mediator paths, where $m_k$ and
$\widetilde m_k$ are the values of the mediator at date $t+k$. Let
$\mathbf w_{-m}$ collect all other values of the endogenous variables through
date $t+r$, which are the same on both paths.

Let $Y_{t+h}\{\mathbf m;\mathbf w_{-m}\}$ denote the value of
the outcome $Y$ at date $t+h$ when the endogenous variables through
date $t+r$ are set to $(\mathbf m,\mathbf w_{-m})$ and the system is
propagated forward from date $t+r$ by \eqref{eq:var}, with the same
coefficients and the same shocks $\varepsilon_{t+r+1},\ldots,\varepsilon_{t+h}$.
Define the controlled mediator-path effect
\begin{align}
\tau_m(h,r;\mathbf m,\widetilde{\mathbf m})
&=
Y_{t+h}\{\mathbf m;\mathbf w_{-m}\}
-
Y_{t+h}\{\widetilde{\mathbf m};\mathbf w_{-m}\}.
\label{eq:controlled_path_effect}
\end{align}

\begin{theorem}[A mediator's contribution as a controlled mediator-path effect]
\label{theolsem}
Suppose Assumption~\ref{ass:svar} holds. For any controlled mediator paths
$\mathbf m$ and $\widetilde{\mathbf m}$,
\begin{align}
\tau_m(h,r;\mathbf m,\widetilde{\mathbf m})
&=
\sum_{k=0}^{r}
\nu_y'
\Phi_{r-k+1}^{(h-r)}
\nu_m
\left(m_k-\widetilde m_k\right).
\label{eq:controlled_path_formula}
\end{align}
In particular, $\tau_m(h,r;\mathbf m,\widetilde{\mathbf m})=\mathcal C_m(h,r)$
if $m_k-\widetilde m_k=\theta_{m,k}$ for $k=0,\ldots,r$.
\end{theorem}

The proof is in Appendix~\ref{prooflsem}.

The shocks after date $t+r$ are the same on both paths and cancel, so
$\tau_m$ depends neither on $\mathbf w_{-m}$ nor on these shocks. A
contribution computed from Sims and higher-order impulse responses is
therefore the effect of an intervention that moves the mediator by its
shock-induced movements $\theta_{m,k}$ and holds the other endogenous variables fixed. Formula \eqref{eq:controlled_path_formula} follows from the VAR alone;
Assumption~\ref{ass:svar} makes $\theta_{m,k}$ the effect of the shock on
the mediator.

At $r=0$, the intervention changes only the value of the mediator at date
$t$. At $r>0$, it changes the values of the mediator at dates
$t,\ldots,t+r$ and holds the other values of the endogenous variables
through date $t+r$ fixed. The effects that earlier movements in the mediator have
on the other variables up to date $t+r$ are then attributed to those
variables. The same mediator can therefore have a large contribution at
early propagation stages and a small contribution at later ones for the
same $h$. Because the other endogenous variables are held fixed, the intervened path
differs from the path the shock generates, in which they also move.

\subsection{Relation to mediation analysis}

Mediation analysis distinguishes controlled, natural and interventional
effects; see \mciteA{imbens2015causal} (\mciteY{imbens2015causal}) for
the potential-outcomes framework and
\mciteA{robins1992identifiability} (\mciteY{robins1992identifiability}),
\mciteA{pearl2022direct} (\mciteY{pearl2022direct}),
\mciteA{vanderweele2015explanation} (\mciteY{vanderweele2015explanation}),
and \mciteA{vansteelandt2017interventional}
(\mciteY{vansteelandt2017interventional}) for mediation. A natural
indirect effect sets the mediator to the value it would take under a
different value of the treatment. \mciteA{imai2010identification}
(\mciteY{imai2010identification}) identify it under sequential
ignorability, which requires the potential outcome under one treatment
value to be independent of the potential mediator under another, given
the treatment and pre-treatment covariates only. This condition compares
different counterfactual regimes, cannot be tested even in a randomized
experiment, and rules out variables that the treatment moves and that
affect both the mediator and the outcome. In a dynamic system with
feedback, such variables are generic. After an oil-price shock, for
example, consumer prices and the federal funds rate can both move and
then affect one another and industrial production, so a variable moved by
the shock affects both the later path of the mediator and the outcome.
This is the standard difficulty with natural effects in longitudinal
settings; see \mciteA{avin2005identifiability}
(\mciteY{avin2005identifiability}) and \mciteA{robins2003semantics}
(\mciteY{robins2003semantics}).

Controlled effects set the mediator to specified values and involve no
potential mediators under other treatment values. Under assumptions of no
unmeasured confounding for the treatment and the mediator, they can be
identified even when the treatment affects variables that later confound
the mediator--outcome relation; see \mciteA{robins1986new}
(\mciteY{robins1986new}), \mciteA{vanderweele2015explanation}
(\mciteY{vanderweele2015explanation}), and \mciteA{acharya2016explaining}
(\mciteY{acharya2016explaining}). The controlled mediator-path effect is
the dynamic counterpart of the controlled effect of a mediator, which
compares the outcome when the mediator is set to two different values
with the treatment held fixed; see \mciteA{vanderweele2011controlled}
(\mciteY{vanderweele2011controlled}). The mediator is set at every date
from $t$ to $t+r$, and the other endogenous variables
through date $t+r$, including those moved by the shock, are held fixed.
A mediator's contribution is therefore a controlled mediator-path effect,
not a natural indirect effect.

\subsection{Relation to counterfactual exercises}
\label{subsec:scenario}

Counterfactual exercises ask how the response would change under an
alternative policy rule or an imposed path. BGW and KL modify the
equation for the federal funds rate, BGW by fixing its path and KL by
removing its direct response to the real price of oil, and implement the
counterfactual with policy shocks. \mciteA{mckay2023can}
(\mciteY{mckay2023can}) construct counterfactual responses under
alternative policy rules from the estimated effects of policy shocks.

Unlike these counterfactual exercises, the decomposition leaves the
estimated law of motion unchanged and allocates a given impulse response across the
endogenous variables at each propagation stage. It requires
identification only of the shock whose impulse response is decomposed; no
policy shock needs to be identified. The contributions of all variables
add up to the impulse response at every propagation stage, whereas the
gaps from separate counterfactual exercises need not add up. The
decomposition also applies to variables that are not policy instruments,
such as consumer prices.

Intervention analysis (\mciteA{box1975intervention},
\mciteY{box1975intervention}), conditional forecasting
(\mciteA{waggoner1999conditional}, \mciteY{waggoner1999conditional}), and
structural scenario analysis (\mciteA{antolin2021structural},
\mciteY{antolin2021structural}) also impose paths on selected variables
and propagate the system forward, and they report the resulting path of
the other variables. We use the same operation for attribution: the two
controlled mediator paths differ by the shock-induced movements in the
mediator, and the resulting difference in the outcome is the mediator's
contribution.

Like these exercises, the controlled mediator-path effect holds the
estimated coefficients fixed when the mediator path changes, so the
critique of \mciteA{lucas1976econometric} (\mciteY{lucas1976econometric})
applies. This assumption is most credible for modest interventions in the
sense of \mciteA{leeper2003modest} (\mciteY{leeper2003modest}). In a
mediator's contribution, the difference between the two mediator paths
is the movement induced by the identified shock. The contribution is
therefore an attribution under the estimated law of motion.

\section{Estimation and inference}
\label{sec:inference}

This section provides estimation and inference for the contributions and
a linear test of the propagation margin.

\subsection{Inference for the contributions}
\label{subsec:channel-inference}

The decomposition does not require a particular estimator. Given a
sample of size $T$, the multi-horizon projection coefficients can be
estimated directly from
\eqref{eq:mhlp_formal} or recursively from an estimated VAR. We first
state the result for recursive VAR estimation, which makes explicit the
dependence of the contributions on the estimated VAR coefficients and
the impact vector. Let
\begin{align}
\xi
&=
\left(
\operatorname{vec}(\Phi_1)',
\ldots,
\operatorname{vec}(\Phi_p)',
\boldsymbol\theta_0'
\right)'.
\label{eq:primitive-parameter}
\end{align}
We use $\widehat{\;\cdot\;}$ to denote estimates. The contribution of
variable $m$ is estimated by
\begin{align}
\widehat{\mathcal C}_m(h,r)
&=
\sum_{k=\max\{0,r-p+1\}}^{r}
\nu_y'
\widehat\Phi_{r-k+1}^{(h-r)}
\nu_m
\,
\widehat\theta_{m,k}.
\label{eq:estimated-channel}
\end{align}
The sum starts at $\max\{0,r-p+1\}$ because
$\Phi_{r-k+1}^{(h-r)}=0$ for $r-k+1>p$. Under recursive estimation,
both $\widehat\Phi_{r-k+1}^{(h-r)}$ and $\widehat\theta_{m,k}$ are
functions of $\widehat\xi$; no further parameters are estimated.

We impose the following high-level condition on the primitive
estimates.

\begin{assumption}[Joint asymptotic normality]
\label{ass:reg}
Under \eqref{eq:var}, suppose that
\begin{align}
\sqrt T
\left(
\widehat\xi-\xi
\right)
&\xrightarrow{d}
N(0,\Omega)
\label{eq:primitive-joint-clt}
\end{align}
and that $\widehat\Omega\xrightarrow{p}\Omega$.
\end{assumption}

Assumption~\ref{ass:reg} lets the estimated impact vector depend on the
same reduced-form estimates as the VAR coefficients, so shock
identification and propagation can be correlated in sampling. The
formulation follows the delta-method argument of
\mciteA{lutkepohl1990asymptotic} (\mciteY{lutkepohl1990asymptotic}) for
impulse responses, but here the mapping also contains the higher-order
impulse responses that enter the decomposition. The contribution
$\mathcal C_m(h,r)$ is a polynomial in the elements of $\xi$. Its
gradient $\nabla_\xi\mathcal C_m(h,r)$ follows from the recursion
\eqref{eq:algrec_recursive} and the product rule and is given in the
proof in Appendix~\ref{proof-channel-asymptotic}. Let
$\widehat\nabla_\xi\mathcal C_m(h,r)$ denote the gradient evaluated at
$\widehat\xi$.

\begin{proposition}[Asymptotic distribution of a contribution]
\label{prop:channel-asymptotic}
Suppose Assumption~\ref{ass:reg} holds. Then
\begin{align}
\sqrt T
\left[
\widehat{\mathcal C}_m(h,r)
-
\mathcal C_m(h,r)
\right]
&\xrightarrow{d}
N\!\left(0,\sigma_m^2(h,r)\right),
\label{eq:channel-delta-limit}
\end{align}
where $\sigma_m^2(h,r)=\nabla_\xi\mathcal C_m(h,r)'\,\Omega\,
\nabla_\xi\mathcal C_m(h,r)$. Moreover,
$\widehat\sigma_m^2(h,r)
=\widehat\nabla_\xi\mathcal C_m(h,r)'\,\widehat\Omega\,
\widehat\nabla_\xi\mathcal C_m(h,r)
\xrightarrow{p}\sigma_m^2(h,r)$.
\end{proposition}

The proof is in Appendix~\ref{proof-channel-asymptotic}. The limit in
\eqref{eq:channel-delta-limit} is degenerate if $\sigma_m^2(h,r)=0$, and
standard first-order normal inference requires $\sigma_m^2(h,r)>0$. If
$\Omega$ is nonsingular, this condition reduces to the rank condition
$\nabla_\xi\mathcal C_m(h,r)\neq0$; \mciteA{inoue2020uniform}
(\mciteY{inoue2020uniform}) impose the same kind of rank condition for
delta-method inference on impulse responses. Under a normalization that
fixes one element of the impact vector, however, $\Omega$ is singular,
and a nonzero gradient is not sufficient.

The derivative of a contribution with respect to the VAR coefficients
has two parts: the derivative of the propagation margin, weighted by the
exposure margin, and the derivative of the exposure margin, weighted by
the propagation margin. The derivative with respect to the impact vector
is weighted by the propagation margin; see
\eqref{eq:contribution-jacobian-var} and
\eqref{eq:contribution-jacobian-impact}. Hence the gradient is zero if,
for every $k$ entering the contribution, both $\theta_{m,k}=0$ and
$\nu_y'\Phi_{r-k+1}^{(h-r)}\nu_m=0$. In that case $\sigma_m^2(h,r)=0$,
and first-order normal inference is degenerate. When both margins are
close to zero, the first-order normal approximation can also be
inaccurate. A zero contribution need not produce this problem: different
summands may offset while $\sigma_m^2(h,r)$ remains positive.
\mciteA{lutkepohl1990asymptotic} (\mciteY{lutkepohl1990asymptotic})
notes the analogous issue for impulse responses whose derivative with
respect to the VAR coefficients vanishes.

A similar rank problem arises in tests of nonlinear Granger
noncausality restrictions; see
\mciteA{boudjellaba1992testing}
(\mciteY{boudjellaba1992testing}). Wald tests require a Jacobian of full
rank, which can fail under the null when nonlinear restrictions are
locally singular; \mciteA{dufour2025wald} (\mciteY{dufour2025wald})
derive the behavior of Wald statistics in this case. We instead test a
sufficient condition for a zero contribution that is linear in the
multi-horizon projection coefficients: a zero propagation margin.

If the multi-horizon projection coefficients are estimated directly
from \eqref{eq:mhlp_formal}, $\xi$ collects the directly estimated
coefficients and the impact vector, each contribution is again a
polynomial in $\xi$, and the same argument applies. A bootstrap provides
an alternative for either estimator. Each replication re-estimates the
VAR or the multi-horizon projections together with the identification of
the shock and recomputes the contributions, which preserves the sampling
dependence between the exposure and propagation margins. With an external
instrument, the instrument is resampled jointly with the reduced-form
observations used to identify the shock.

\subsection{Testing the propagation margin}
\label{subsec:linear-channel-test}

The null hypothesis that the
channel through $m$ is closed at $(h,r)$ is
\begin{align}
\mathcal H_0^{\mathrm P}:
\qquad
\nu_y'\Phi_k^{(h-r)}\nu_m
&=
0,
\qquad
k=1,\ldots,\min\{r+1,p\}.
\label{eq:channel_linear_null}
\end{align}
The coefficient on lag $k$ is the propagation margin of the movement in
variable $m$ at intermediate horizon $r-k+1$. The propagation margins at
intermediate horizons before $r-p+1$ are zero because $\Phi_k^{(h-r)}=0$ for $k>p$. Under
$\mathcal H_0^{\mathrm P}$, $\mathcal C_m(h,r)=0$ for every
shock. For $k\geq2$, the coefficient $\nu_y'\Phi_k^{(h-r)}\nu_m$ is the response
of the best linear forecast of $Y_{t+h-r}$ to variable $m$ at date
$t+1-k$. The regressors held fixed then include the values of variable $m$
at the later dates $t+2-k,\ldots,t$, so a higher value at date $t+1-k$
means that variable $m$ moved earlier. These
coefficients therefore show whether the timing of a movement in the
mediator matters for the outcome, given its value at date $t$.
For consumer prices, for example, they show whether industrial production
$h-r$ months ahead depends on when consumer prices rose, given their values
at the later dates.

The hypothesis $\mathcal H_0^{\mathrm P}$ states that movements in
variable $m$ at the intermediate horizons that enter the contribution at propagation stage
$r$ do not carry over to the outcome at horizon $h$. Rejecting $\mathcal H_0^{\mathrm P}$ shows that the channel through $m$ is
open at $(h,r)$; failing to reject it does not show that the channel is
closed.

Although the contributions can be
computed from an estimated VAR, it is more convenient to test
$\mathcal H_0^{\mathrm P}$ by estimating \eqref{eq:mhlp_formal} directly
by OLS, regressing $Y_{t+h-r}$ on
$W_t,\ldots,W_{t-p+1}$. The hypothesis $\mathcal H_0^{\mathrm P}$ is then
a set of linear restrictions on regression coefficients and can be tested
by a joint Wald test; see \mciteA{dufour2006short}
(\mciteY{dufour2006short}). The error in this regression is a moving
average of order $h-r-1$, so the Wald test uses a
heteroskedasticity- and autocorrelation-consistent (HAC) covariance
matrix. The number of restrictions, $\min\{r+1,p\}$, increases with $r$ until
$r=p-1$: up to that stage, each later propagation stage adds a movement
in the mediator whose propagation margin the null restricts.

Multi-horizon Granger noncausality gives a sufficient condition for
$\mathcal H_0^{\mathrm P}$. Following \mciteA{dufour1998short}
(\mciteY{dufour1998short}), variable $m$ does not Granger-cause $Y$ at
horizon $h-r$ if removing the past of $m$ from the information set does
not change the best linear forecast of $Y_{t+h-r}$; we write
$m\overset{h-r}{\nrightarrow}Y$. In a VAR of order $p$,
\begin{align}
m\overset{h-r}{\nrightarrow}Y
\quad\Longleftrightarrow\quad
\nu_y'\Phi_k^{(h-r)}\nu_m=0,
\qquad
k=1,\ldots,p.
\label{eq:mhgnc}
\end{align}
The restrictions in \eqref{eq:mhgnc} contain those in
\eqref{eq:channel_linear_null}. Multi-horizon Granger noncausality from
$m$ to $Y$ at horizon $h-r$ therefore implies that the channel through $m$
is closed at $(h,r)$ and that $\mathcal C_m(h,r)=0$. When
$r+1\geq p$, the two sets of restrictions coincide. When $r+1<p$,
$\mathcal H_0^{\mathrm P}$ imposes only the $r+1$ restrictions that enter
the contribution at propagation stage $r$. Because multi-horizon Granger
causality concerns only the propagation margin, a mediator can
Granger-cause the outcome at every horizon and still make a small
contribution to the impulse response of the outcome to a particular
shock.

\section{Empirical application: transmission of oil-price shocks}
\label{sec:app}

We decompose the impulse response of industrial production to an oil
supply news shock. The two mediators are the federal funds rate, which is
central to the BGW hypothesis, and consumer prices, which capture the
pass-through of higher oil prices to the prices faced by households. We
ask how much the movements in each
mediator contribute to the decline in industrial production and at which
propagation stages these contributions arise. Whether movements in the
federal funds rate affect output is a question about the propagation
margin, which we test separately.

The VAR has five variables: the real commodity price, the real price of
oil, industrial production, consumer prices and the federal funds rate.
It has twelve lags and is estimated on monthly U.S.\ data from 1975:01 to
2019:12. All variables except the federal funds rate enter in logs. The
variables follow BGW as modified by KL, except that we use industrial
production instead of a coincident index, so that the outcome is a
directly observed series. With the external-instrument identification
used here, the contributions do not depend on how the variables are
ordered.

The oil supply news shock is identified with the external instrument of
\mciteA{kanzig2021macroeconomic} (\mciteY{kanzig2021macroeconomic}), the
high-frequency revision in oil futures prices around OPEC production
announcements, following \mciteA{mertens2013dynamic}
(\mciteY{mertens2013dynamic}). The instrument is available from 1983:04
to 2019:12 and is nonzero in $121$ months. Its correlation with the
reduced-form residual of the oil price is $0.19$, and the first-stage $F$
statistic is $16.56$ ($p=5.6\times10^{-5}$). The identification sample
thus starts after the Volcker disinflation. It covers the second
subsample of KL and extends it by a decade.

We normalize the shock so
that the real price of oil rises by ten percent on impact. On impact,
commodity prices rise by $1.63$ percent, industrial production falls by
$0.21$ percent, consumer prices rise by $0.14$ percent and the federal
funds rate rises by $2.3$ basis points. In this section, the outcome $Y$
is industrial production.

\begin{figure}[!htbp]
\centering
\caption[Exposure margins]{}
\label{fig:irf}
\begin{center}
Figure \thefigure. Impulse responses to a 10\% oil supply news shock
\end{center}
\includegraphics[width=\textwidth]{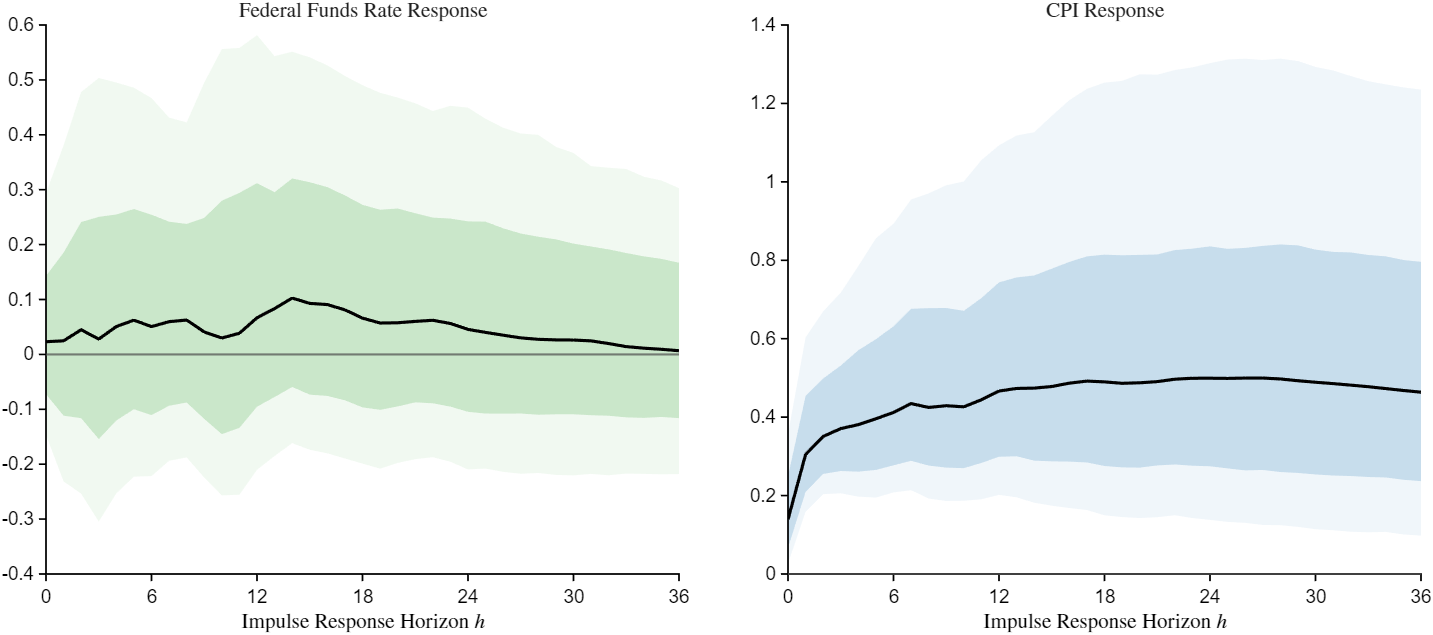}

\begin{minipage}{\textwidth}
\footnotesize
\vspace{0.5em}
\emph{Note.} Impulse responses of the
federal funds rate and consumer prices to an oil supply news shock
identified with an external instrument and normalized to raise the real
price of oil by ten percent on impact. The federal funds rate is measured
in percentage points and consumer prices in log points. Shaded areas are
$68$ and $90$ percent moving-block bootstrap confidence bands (block
length $24$ months, $2{,}000$ draws). Reduced-form sample
1975:01--2019:12; identification sample 1983:04--2019:12.
\end{minipage}
\end{figure}

\begin{figure}[!htbp]
\centering
\caption[Dynamic decomposition of industrial production]{}
\label{fig:ipdecomp}
\begin{center}
Figure \thefigure. Dynamic decomposition of the impulse response of
industrial production across propagation stages
\end{center}
\includegraphics[width=\textwidth]{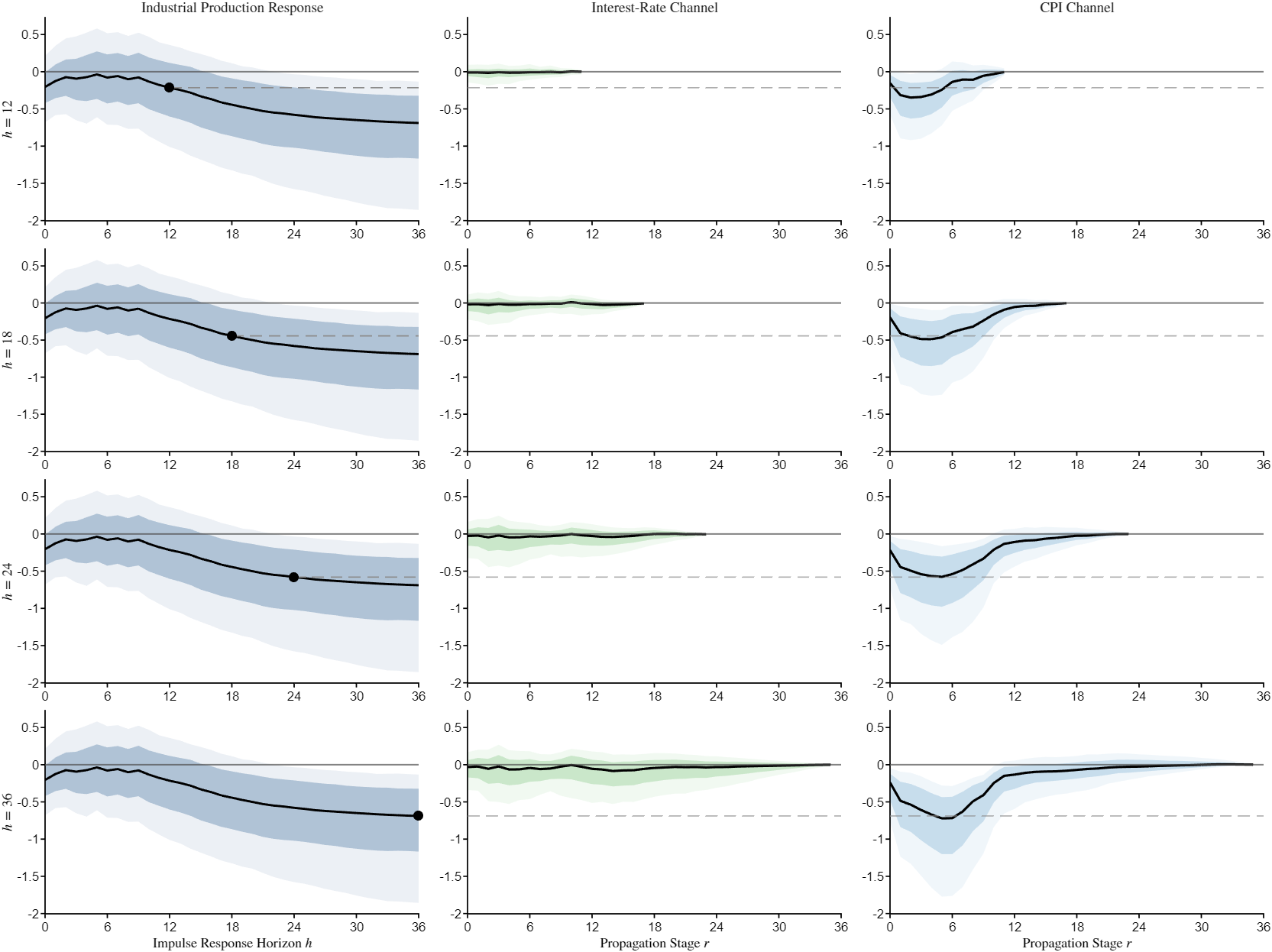}

\begin{minipage}{\textwidth}
\footnotesize
\vspace{0.5em}
\emph{Note.} Each row decomposes the impulse response of industrial
production at a specific horizon $h$ of interest, with $h=12$, $18$, $24$
and $36$ months in the four rows. The left panel shows
the impulse response of industrial production; the dot marks its value
$\theta_{y,h}$ at horizon $h$, which is the value being decomposed, and the dashed line carries this value across the row. The
center and right panels show the contributions $\mathcal C_m(h,r)$ of the
federal funds rate and consumer prices to this value, plotted against the
propagation stage $r=0,\ldots,h-1$. At every $r$, these two contributions and those of the
other three variables sum to the dashed line. Contributions at different
propagation stages should not be added. Shaded areas are $68$ and $90$ percent moving-block bootstrap
bands.
\end{minipage}
\end{figure}

We compute the impulse responses and the contributions from the
estimated VAR. To test $\mathcal H_0^{\mathrm P}$, we estimate the
multi-horizon projection coefficients by OLS instead, with HAC standard
errors. For the confidence bands in Figures~\ref{fig:irf} and \ref{fig:ipdecomp},
we use the moving-block bootstrap of \mciteA{jentsch2019asymptotically}
(\mciteY{jentsch2019asymptotically}) with a block length of $24$ months
and $2{,}000$ draws; each draw re-estimates the VAR and re-identifies the
shock with the instrument.

Figure~\ref{fig:irf} reports the impulse responses of the two mediators,
that is, their exposure margins. The impulse response of the federal
funds rate is positive but small. It peaks at ten basis points at $h=14$,
falls below five basis points by $h=24$ and is close to zero at $h=36$.
The $68$ percent band contains zero at every horizon. KL likewise find
little evidence after 1987 that oil-price shocks move
the federal funds rate systematically. The small response is also
consistent with the supply-side nature of the shock, which raises prices
while depressing real activity and thus confronts monetary policy with a
trade-off between inflation and activity.

Consumer prices rise by $0.14$ percent on impact, by $0.41$ percent within
six months and by $0.50$ percent within two years, and the $68$ percent
band excludes
zero at every horizon after impact. The exposure margin of consumer
prices is large and builds up within six months, and the
exposure margin of the federal funds rate is small.

The impulse response of industrial production is $-0.21$ percent on
impact, is flat through the first nine months and then falls steadily to
$-0.69$ percent at $h=36$. Figure~\ref{fig:ipdecomp} decomposes its values at $h=12$, $18$, $24$
and $36$ months at every propagation stage $r<h$. The vertical distance between the sum of the two contributions
shown and the dashed line is the sum of the contributions of the other
three variables at that propagation stage.

The contribution of the federal funds rate is small at each of the four horizons and at every propagation stage. The largest contribution
in the figure is $-0.085$ percentage points, at $h=36$ and $r=14$,
against an impulse response of industrial production of $-0.689$ percent.
The $68$ percent band contains zero at every pair $(h,r)$. The maximum
occurs at $r=14$, the horizon at which the
impulse response of the federal funds rate peaks, $\theta_{\mathrm{ffr},k}$
being largest at $k=14$. KL find that their counterfactual impulse response of output differs
little from the estimated one. The contribution of the federal funds rate
is a different object, but its small size at every horizon and propagation stage we consider is consistent with that finding.

The contribution of consumer prices is large already at early
propagation stages. At $h=12$, the impulse response of industrial
production is $-0.215$ percent, close to its value on impact, yet the rise
in consumer prices through month two contributes $-0.347$ percentage
points. At the three longer horizons, the contribution of
consumer prices at its peak over $r$ is $-0.487$ at $h=18$, $-0.576$ at
$h=24$ and $-0.722$ at $h=36$. The corresponding impulse responses of
industrial production are $-0.443$, $-0.579$ and $-0.689$ percent. The
$68$ percent
band lies below zero at the propagation stages where the contribution is
largest.

At $(h,r)=(12,2)$, $(18,4)$ and $(36,5)$, the contribution of consumer
prices is larger in absolute value than the impulse response of
industrial production. The contributions of the other three variables,
the real price of oil, the real commodity price and industrial production
itself, are positive in sum and offset part of the contribution of
consumer prices, as in the numerical example of
Section~\ref{subsec:num_illustration}.

The contribution of consumer prices peaks at $r=2$ for $h=12$ and at
$r=4$ or $5$ for the three longer horizons. It falls below half
its peak by $r\approx6$ at $h=12$ and by $r\approx10$ at $h=24$ and
$h=36$. At $h=36$, it is below a fifth of its peak by $r=12$ and below a
tenth by $r=20$. At each of the four horizons, the movements in consumer prices
through month six account for much of the decline in industrial
production. At later propagation stages, most of the impulse response
is accounted for by the other endogenous variables, whose
movements by then include the effects of the earlier rise in consumer
prices.

\begin{sidewaystable}[p]
\centering
\caption{Tests of the propagation margin for the impulse response of
industrial production}
\label{tab:lintests}
\begin{threeparttable}
\scriptsize
\setlength{\tabcolsep}{8pt}
\begin{tabular}{cc|rrrrrrrrrrrr|c}
\hline\hline
$h$ & $r$
& $k=1$ & $2$ & $3$ & $4$ & $5$ & $6$ & $7$ & $8$ & $9$ & $10$ & $11$
& $12$ & Joint $p$ \\
\hline
\multicolumn{15}{l}{\emph{Panel A. Mediator: federal funds rate}}\\[2pt]
12 & 0  & $-0.470$ & & & & & & & & & & & & 0.249 \\
12 & 3  & $-0.384$ & $-0.083$ & $0.238$ & $0.142$ & & & & & & & & & 0.341 \\
12 & 6  & $-0.415$ & $0.023$ & $0.027$ & $0.328$ & $-0.086$ & $0.235$ & $0.044$ & & & & & & 0.727 \\
12 & 11 & $0.104$ & $-0.162$ & $-0.070$ & $0.096$ & $-0.029$ & $0.014$ & $0.090$ & $-0.079$ & $0.108$ & $-0.195$ & $0.198$ & $-0.101$ & 0.114 \\[3pt]
18 & 0  & $-0.523$ & & & & & & & & & & & & 0.214 \\
18 & 6  & $-0.470$ & $-0.023$ & $0.196$ & $0.236$ & $0.084$ & $0.092$ & $0.197$ & & & & & & 0.481 \\
18 & 12 & $-0.415$ & $0.023$ & $0.027$ & $0.328$ & $-0.086$ & $0.235$ & $0.044$ & $-0.232$ & $0.180$ & $-0.106$ & $0.176$ & $-0.397$ & 0.659 \\
18 & 17 & $0.104$ & $-0.162$ & $-0.070$ & $0.096$ & $-0.029$ & $0.014$ & $0.090$ & $-0.079$ & $0.108$ & $-0.195$ & $0.198$ & $-0.101$ & 0.114 \\[3pt]
24 & 0  & $-0.959^{**}$ & & & & & & & & & & & & $0.023^{**}$ \\
24 & 8  & $-0.418$ & $-0.030$ & $0.129$ & $0.027$ & $0.142$ & $-0.132$ & $0.199$ & $-0.184$ & $0.163$ & & & & 0.829 \\
24 & 16 & $-0.438$ & $0.035$ & $0.070$ & $0.316$ & $-0.066$ & $0.129$ & $0.205$ & $-0.170$ & $0.129$ & $-0.180$ & $0.180$ & $-0.534$ & 0.588 \\
24 & 23 & $0.104$ & $-0.162$ & $-0.070$ & $0.096$ & $-0.029$ & $0.014$ & $0.090$ & $-0.079$ & $0.108$ & $-0.195$ & $0.198$ & $-0.101$ & 0.114 \\[3pt]
36 & 0  & $-1.336^{**}$ & & & & & & & & & & & & $0.004^{**}$ \\
36 & 12 & $-0.959^{**}$ & $0.182$ & $-0.014$ & $-0.023$ & $0.051$ & $0.014$ & $0.141$ & $-0.341^{**}$ & $0.124$ & $0.052$ & $-0.085$ & $-0.203$ & $0.001^{**}$ \\
36 & 24 & $-0.470$ & $-0.023$ & $0.196$ & $0.236$ & $0.084$ & $0.092$ & $0.197$ & $-0.319$ & $0.260$ & $-0.405^{*}$ & $0.163$ & $-0.559$ & $0.044^{**}$ \\
36 & 35 & $0.104$ & $-0.162$ & $-0.070$ & $0.096$ & $-0.029$ & $0.014$ & $0.090$ & $-0.079$ & $0.108$ & $-0.195$ & $0.198$ & $-0.101$ & 0.114 \\
\hline
\multicolumn{15}{l}{\emph{Panel B. Mediator: consumer prices}}\\[2pt]
12 & 0  & $-2.002^{**}$ & & & & & & & & & & & & $0.013^{**}$ \\
12 & 3  & $-1.098$ & $-0.233$ & $0.172$ & $-0.292$ & & & & & & & & & $0.038^{**}$ \\
12 & 6  & $-0.588$ & $-0.142$ & $0.080$ & $-0.094$ & $-0.057$ & $0.340$ & $0.257$ & & & & & & 0.378 \\
12 & 11 & $0.162$ & $-0.319$ & $0.182$ & $-0.234$ & $0.256$ & $-0.290$ & $0.407^{*}$ & $-0.238$ & $0.019$ & $-0.049$ & $0.089$ & $0.015$ & 0.639 \\[3pt]
18 & 0  & $-2.165^{**}$ & & & & & & & & & & & & $0.014^{**}$ \\
18 & 6  & $-2.002^{**}$ & $0.301$ & $0.126$ & $-0.303$ & $0.358$ & $0.529$ & $0.411$ & & & & & & $0.037^{**}$ \\
18 & 12 & $-0.588$ & $-0.142$ & $0.080$ & $-0.094$ & $-0.057$ & $0.340$ & $0.257$ & $-0.269$ & $0.179$ & $0.091$ & $0.403$ & $-0.208$ & 0.552 \\
18 & 17 & $0.162$ & $-0.319$ & $0.182$ & $-0.234$ & $0.256$ & $-0.290$ & $0.407^{*}$ & $-0.238$ & $0.019$ & $-0.049$ & $0.089$ & $0.015$ & 0.639 \\[3pt]
24 & 0  & $-2.101^{**}$ & & & & & & & & & & & & $0.032^{**}$ \\
24 & 8  & $-2.275^{**}$ & $0.772$ & $-0.114$ & $-0.174$ & $0.390$ & $0.207$ & $0.572$ & $0.193$ & $0.192$ & & & & 0.351 \\
24 & 16 & $-0.919$ & $-0.080$ & $-0.153$ & $0.042$ & $-0.027$ & $0.292$ & $0.519$ & $-0.390$ & $0.298$ & $0.470$ & $0.488$ & $-0.551$ & $0.088^{*}$ \\
24 & 23 & $0.162$ & $-0.319$ & $0.182$ & $-0.234$ & $0.256$ & $-0.290$ & $0.407^{*}$ & $-0.238$ & $0.019$ & $-0.049$ & $0.089$ & $0.015$ & 0.639 \\[3pt]
36 & 0  & $-1.256$ & & & & & & & & & & & & 0.382 \\
36 & 12 & $-2.101^{**}$ & $0.427$ & $0.189$ & $0.126$ & $0.092$ & $0.290$ & $0.410$ & $0.511$ & $0.238$ & $0.124$ & $1.258^{*}$ & $-1.491^{**}$ & 0.218 \\
36 & 24 & $-2.002^{**}$ & $0.301$ & $0.126$ & $-0.303$ & $0.358$ & $0.529$ & $0.411$ & $0.033$ & $0.405$ & $0.147$ & $1.028^{**}$ & $-1.039^{*}$ & $0.040^{**}$ \\
36 & 35 & $0.162$ & $-0.319$ & $0.182$ & $-0.234$ & $0.256$ & $-0.290$ & $0.407^{*}$ & $-0.238$ & $0.019$ & $-0.049$ & $0.089$ & $0.015$ & 0.639 \\
\hline\hline
\end{tabular}
\begin{tablenotes}[flushleft]
\footnotesize
\item \emph{Note.}
Entries are OLS estimates of the multi-horizon projection coefficients
$\nu_y'\Phi_k^{(h-r)}\nu_m$ in $\mathcal H_0^{\mathrm P}$, where the
outcome $Y$ is industrial production and $k$ is the lag. Blank cells are lags
that do not enter the contribution at propagation stage $r$; the number
of restrictions is $\min\{r+1,p\}$, with $p=12$. The coefficients depend
on $(h,r)$ only through $h-r$; the joint $p$-values differ because the number of
restrictions differs. When $r\geq p-1$, $\mathcal H_0^{\mathrm P}$
includes all $p$ lags and is multi-horizon Granger noncausality from $m$
to industrial production at horizon $h-r$. The last column reports the
$p$-value of the joint Wald test of $\mathcal H_0^{\mathrm P}$. Joint $p$-values and the
stars on individual coefficients are based on HAC standard errors: $^{*}$
$p<0.10$, $^{**}$ $p<0.05$. 
\end{tablenotes}
\end{threeparttable}
\end{sidewaystable}

Table~\ref{tab:lintests} reports tests of
$\mathcal H_0^{\mathrm P}$ for both mediators at the four horizons and at four propagation stages
for each. Among the reported specifications in Panel A, the test rejects when
$h-r$ is $24$ or $36$ and in the row $(h,r)=(36,24)$. In the row
$(h,r)=(36,0)$, the leading coefficient is $-1.34$ and the single
restriction is rejected with $p=0.004$. Holding the other regressors
fixed, a federal funds rate one percentage point higher at date $t$
lowers the best linear forecast of industrial production three years
ahead by $1.34$ percent. In the row $(24,0)$, the leading coefficient is
$-0.96$ with $p=0.023$. The row $(36,12)$ has the same $h-r=24$ and all
twelve lags, so it is a test of multi-horizon Granger noncausality from
the federal funds rate to industrial production at two years; it rejects
with $p=0.001$. The row $(36,24)$, with $h-r=12$ and all twelve
lags, rejects with $p=0.044$. The negative coefficients are consistent with conventional monetary
transmission.

The tests reject a closed federal funds rate channel at several pairs
$(h,r)$, most of them with long remaining horizons
$h-r$, while the exposure margin of the
federal funds rate is small and imprecisely estimated at every $k$. In
these configurations the channel is open but largely unused by the oil
supply news shock, and the contribution of the federal funds rate is small
because the shock moves the federal funds rate little. The timing of the two margins also keeps the contribution
small. The
leading coefficient of the federal funds rate is between $-0.38$ and
$-0.52$ when $h-r$ is between $6$ and $18$ months and rises in absolute
value to $-0.96$ at $h-r=24$ and $-1.34$ at $h-r=36$. The largest leading
coefficients thus apply to movements in the federal funds rate at early
propagation stages, before its impulse response peaks. The propagation
stages that include the peak, $r\geq14$, have $h-r\leq22$ at $h=36$,
shorter than the remaining horizons with the largest leading
coefficients.

In Panel B, the leading coefficient is between $-2.0$ and $-2.3$ when
$h-r$ is $12$, $16$, $18$ or $24$, with marginal $p$-values of at most $0.032$.
Holding the other regressors fixed, including the real price of oil and
commodity prices, consumer prices one percent higher at date $t$ lower the
best linear forecast of industrial production sixteen months ahead by
$2.3$ percent. Unlike
that of the federal funds rate, the leading coefficient of consumer prices
is hump-shaped in $h-r$. It peaks near $h-r=16$ and declines to $-1.26$ at
$h-r=36$. For consumer prices, both margins are large, and so is their
contribution. This is consistent with the demand-side mechanism emphasized by KL and
\mciteA{edelstein2009sensitive} (\mciteY{edelstein2009sensitive}), in
which energy-price increases reduce household purchasing power and lower
expenditures, in particular on energy-using durables; see also
\mciteA{baumeister2018discretionary} (\mciteY{baumeister2018discretionary}).
The VAR contains no measure of household expenditures, however, so the
decomposition does not isolate this mechanism. In the rows with $r=h-1$, the test does not
reject for either mediator, and at this last propagation stage the
contributions of both mediators are close to zero at each of the four
horizons.

The joint $p$-values need not track the significance of the leading
coefficient. As $r$ increases, the null of a closed channel also restricts
the propagation margins of later movements in the mediator, so the joint
test concerns a larger set of coefficients. The rows $(h,r)=(24,0)$ and
$(36,12)$ of Panel B have the same $h-r=24$ and hence the same leading
coefficient, $-2.10$, with a marginal $p$-value of $0.032$. With one
restriction, the joint $p$-value is $0.032$; with twelve, it is $0.218$.
Similarly, in the row $(24,8)$ the leading coefficient is $-2.28$ with a
marginal $p$-value of $0.006$, while the joint $p$-value over nine
restrictions is $0.351$. A significant leading coefficient therefore need
not lead to rejection of the closed-channel null at later propagation
stages. At small $r$, the null restricts only the propagation margins of the early
movements in consumer prices.

\section{Conclusion}
\label{sec:con}

This paper decomposes the impulse response to a macroeconomic shock into
the contributions of mediator variables. Using the recursion between Sims
and higher-order impulse responses of \mciteA{dufour1998short}
(\mciteY{dufour1998short}), we write the impulse response at a given horizon in terms of the shock-induced movements in the endogenous variables up to
any propagation stage. A mediator's contribution
combines an exposure margin
and a propagation margin. The propagation margin relies on the higher-order impulse responses, the
coefficients on the lags in the multi-horizon projections, which
applications rarely report. A mediator's contribution equals a controlled
mediator-path effect, which gives it a causal interpretation without the
cross-world assumptions of natural indirect effects. Multi-horizon
Granger noncausality sets the propagation margin to zero, and a linear
test of the propagation margin complements inference on the
contributions.

For U.S.\ oil supply news shocks, the contribution of the federal funds
rate to the decline in industrial production is small at every horizon and propagation stage we consider. The tests reject a closed channel through the federal funds rate at
several pairs of horizon and propagation stage, but the shock
moves the federal funds rate too little to generate a large contribution. Consumer prices have large exposure and propagation margins, and
their movements in the first six months account for most of the later
decline in industrial production.
The decomposition is linear and, like the impulse response it decomposes,
depends on the variables included in the VAR. Extending it to nonlinear or state-dependent systems
would let both margins vary with economic conditions.

\appendix

\section{Proof of Proposition~\ref{prop:iterated_phi_identity}}
\label{proofprop}

\begin{proof}
We argue by induction on $r$. For $r=0$, the right-hand side of
\eqref{irfhn} is $\Phi_1^{(h)}\Phi_1^{(0)}=\Phi_1^{(h)}$. Suppose that
\eqref{irfhn} holds for some $r\leq h-2$. Because $h-r-1\geq1$, the
recursion \eqref{eq:algrec_recursive} at horizon $h-r-1$ gives, for
$k=0,\ldots,r$,
\begin{align}
\Phi_{r-k+1}^{(h-r)}
&=
\Phi_{r-k+2}^{(h-r-1)}
+
\Phi_1^{(h-r-1)}\Phi_{r-k+1}.
\label{eq:proof_step}
\end{align}
By \eqref{eq:sims_direct_mediated} with $h$ replaced by $r$, and
$\Phi_1^{(0)}=I$,
\begin{align}
\sum_{k=0}^{r}\Phi_{r-k+1}\Phi_1^{(k)}
&=
\Phi_{r+1}
+
\sum_{k=1}^{r}\Phi_{r-k+1}\Phi_1^{(k)}
=
\Phi_1^{(r+1)}.
\label{eq:proof_sims}
\end{align}
Substituting \eqref{eq:proof_step} into \eqref{irfhn} and using
\eqref{eq:proof_sims},
\begin{align}
\begin{split}
\Phi_1^{(h)}
&=
\sum_{k=0}^{r}\Phi_{r-k+2}^{(h-r-1)}\Phi_1^{(k)}
+
\Phi_1^{(h-r-1)}\sum_{k=0}^{r}\Phi_{r-k+1}\Phi_1^{(k)}
\\
&=
\sum_{k=0}^{r}\Phi_{r-k+2}^{(h-r-1)}\Phi_1^{(k)}
+
\Phi_1^{(h-r-1)}\Phi_1^{(r+1)}
\\
&=
\sum_{k=0}^{r+1}\Phi_{(r+1)-k+1}^{(h-(r+1))}\Phi_1^{(k)},
\end{split}
\label{eq:proof_induction}
\end{align}
which is \eqref{irfhn} with $r$ replaced by $r+1$. Hence \eqref{irfhn}
holds for all $0\leq r<h$. Post-multiplying \eqref{irfhn} by
$\boldsymbol\theta_0$ and using $\boldsymbol\theta_k=\Phi_1^{(k)}\boldsymbol\theta_0$
for $k\geq0$ gives \eqref{eq:theta_recursion}.
\end{proof}

\section{Proof of Theorem~\ref{theolsem}}
\label{prooflsem}

\begin{proof}
Let $W^{*}_s$ denote the vector of endogenous variables at date $s\leq t+r$
when the endogenous variables through date $t+r$ are set to
$(\mathbf m,\mathbf w_{-m})$,
and let $\widetilde W^{*}_s$ denote it when it is set to
$(\widetilde{\mathbf m},\mathbf w_{-m})$. After date $t+r$, both
controlled paths follow \eqref{eq:var} with the same coefficients and the
same shocks. Iterating \eqref{eq:var} forward from date $t+r$, as in the
derivation of \eqref{eq:mhlp_formal}, and premultiplying by $\nu_y'$
gives
\begin{align}
Y_{t+h}\{\mathbf m;\mathbf w_{-m}\}
&=
\sum_{\ell=1}^{p}\nu_y'\Phi_\ell^{(h-r)}W^{*}_{t+r+1-\ell}
+
\nu_y'\sum_{q=0}^{h-r-1}\Psi_q\Theta_0\varepsilon_{t+h-q},
\label{eq:proof_potential}
\end{align}
and the same expression with $\widetilde W^{*}$ in place of $W^{*}$ for
$Y_{t+h}\{\widetilde{\mathbf m};\mathbf w_{-m}\}$. Subtracting the two
expressions eliminates the last term. The two paths differ only in
variable $m$ at dates $t,\ldots,t+r$:
\begin{align}
W^{*}_{t+k}-\widetilde W^{*}_{t+k}
&=
\begin{cases}
\nu_m\left(m_k-\widetilde m_k\right), & k=0,\ldots,r,\\
0, & k<0.
\end{cases}
\label{eq:proof_path_difference}
\end{align}
Date $t+r+1-\ell$ is date $t+k$ with $k=r+1-\ell$, so
\begin{align}
\begin{split}
\tau_m(h,r;\mathbf m,\widetilde{\mathbf m})
&=
\sum_{\ell=1}^{p}\nu_y'\Phi_\ell^{(h-r)}
\left(W^{*}_{t+r+1-\ell}-\widetilde W^{*}_{t+r+1-\ell}\right)
\\
&=
\sum_{k=\max\{0,r+1-p\}}^{r}
\nu_y'\Phi_{r-k+1}^{(h-r)}\nu_m\left(m_k-\widetilde m_k\right)
\\
&=
\sum_{k=0}^{r}
\nu_y'\Phi_{r-k+1}^{(h-r)}\nu_m\left(m_k-\widetilde m_k\right),
\end{split}
\label{eq:proof_tau}
\end{align}
where the last equality uses $\Phi_\ell^{(h-r)}=0$ for $\ell>p$. This is
\eqref{eq:controlled_path_formula}, and it contains neither
$\mathbf w_{-m}$ nor the shocks. Setting
$m_k-\widetilde m_k=\theta_{m,k}$ for $k=0,\ldots,r$ gives
$\mathcal C_m(h,r)$ in \eqref{eq:local_channel}.
\end{proof}

\section{Proof of Proposition~\ref{prop:channel-asymptotic}}
\label{proof-channel-asymptotic}

\begin{proof}
The contribution $\mathcal C_m(h,r)$ is a polynomial in the elements of
$\xi$: by \eqref{eq:algrec_recursive}, the multi-horizon projection
coefficients are polynomials in the VAR coefficients, and
$\theta_{m,k}=\nu_m'\Phi_1^{(k)}\boldsymbol\theta_0$. Write
$\mathcal C_m(h,r)=g(\xi)$, where $g$ is continuously differentiable with
gradient $\nabla_\xi\mathcal C_m(h,r)$. By Assumption~\ref{ass:reg}
and the delta method,
\begin{align}
\begin{split}
\sqrt T\left[\widehat{\mathcal C}_m(h,r)-\mathcal C_m(h,r)\right]
&=
\nabla_\xi\mathcal C_m(h,r)'\,\sqrt T\left(\widehat\xi-\xi\right)+o_p(1)
\\
&\xrightarrow{d}
N\!\left(0,\sigma_m^2(h,r)\right).
\end{split}
\label{eq:proof_delta}
\end{align}
Finally, Assumption~\ref{ass:reg} implies
$\widehat\xi\xrightarrow{p}\xi$, so
$\widehat\nabla_\xi\mathcal C_m(h,r)\xrightarrow{p}
\nabla_\xi\mathcal C_m(h,r)$ by continuity; together with
$\widehat\Omega\xrightarrow{p}\Omega$, this gives
$\widehat\sigma_m^2(h,r)\xrightarrow{p}\sigma_m^2(h,r)$.

It remains to compute the gradient. For $\ell=1,\ldots,p$,
$k=1,\ldots,p$ and $q\geq1$, differentiating
\eqref{eq:algrec_recursive} gives
\begin{align}
\begin{split}
\frac{\partial
\operatorname{vec}\!\left(\Phi_k^{(q+1)}\right)}
{\partial\operatorname{vec}(\Phi_\ell)'}
&=
\frac{\partial
\operatorname{vec}\!\left(\Phi_{k+1}^{(q)}\right)}
{\partial\operatorname{vec}(\Phi_\ell)'}
+
\left(\Phi_k'\otimes I_K\right)
\frac{\partial
\operatorname{vec}\!\left(\Phi_1^{(q)}\right)}
{\partial\operatorname{vec}(\Phi_\ell)'}
\\
&\quad+
\mathbf 1\{k=\ell\}
\left(I_K\otimes\Phi_1^{(q)}\right),
\end{split}
\label{eq:mhlp-jacobian-recursion}
\end{align}
where the first term is zero for $k=p$ because $\Phi_{p+1}^{(q)}=0$.
Starting from $\partial\operatorname{vec}(\Phi_k^{(1)})/
\partial\operatorname{vec}(\Phi_\ell)'=\mathbf 1\{k=\ell\}I_{K^2}$,
this recursion gives the derivatives of the Sims and higher-order
impulse-response matrices. Because
$\theta_{m,k}=\nu_m'\Phi_1^{(k)}\boldsymbol\theta_0$ for $k\geq1$,
\begin{align}
\begin{split}
\frac{\partial\theta_{m,k}}
{\partial\operatorname{vec}(\Phi_\ell)'}
&=
\left(
\boldsymbol\theta_0'\otimes\nu_m'
\right)
\frac{\partial
\operatorname{vec}\!\left(\Phi_1^{(k)}\right)}
{\partial\operatorname{vec}(\Phi_\ell)'},
\\[6pt]
\frac{\partial\theta_{m,k}}
{\partial\boldsymbol\theta_0'}
&=
\nu_m'\Phi_1^{(k)}.
\end{split}
\label{eq:exposure-jacobian}
\end{align}
For $k=0$, $\theta_{m,0}=\nu_m'\boldsymbol\theta_0$, so the derivative
is zero with respect to the VAR coefficients and $\nu_m'$ with respect to
$\boldsymbol\theta_0$. By the product rule, for $\ell=1,\ldots,p$,
\begin{align}
\begin{split}
\frac{\partial\mathcal C_m(h,r)}
{\partial\operatorname{vec}(\Phi_\ell)'}
&=
\sum_{k=\max\{0,r-p+1\}}^{r}
\theta_{m,k}
\left(
\nu_m'\otimes\nu_y'
\right)
\frac{\partial
\operatorname{vec}\!\left(
\Phi_{r-k+1}^{(h-r)}
\right)}
{\partial\operatorname{vec}(\Phi_\ell)'}
\\
&\quad+
\sum_{k=\max\{1,r-p+1\}}^{r}
\left(
\nu_y'
\Phi_{r-k+1}^{(h-r)}
\nu_m
\right)
\left(
\boldsymbol\theta_0'\otimes\nu_m'
\right)
\frac{\partial
\operatorname{vec}\!\left(\Phi_1^{(k)}\right)}
{\partial\operatorname{vec}(\Phi_\ell)'},
\end{split}
\label{eq:contribution-jacobian-var}
\end{align}
and
\begin{align}
\frac{\partial\mathcal C_m(h,r)}
{\partial\boldsymbol\theta_0'}
&=
\sum_{k=\max\{0,r-p+1\}}^{r}
\left(
\nu_y'
\Phi_{r-k+1}^{(h-r)}
\nu_m
\right)
\nu_m'\Phi_1^{(k)},
\label{eq:contribution-jacobian-impact}
\end{align}
with $\Phi_1^{(0)}=I$. Stacking \eqref{eq:contribution-jacobian-var}
for $\ell=1,\ldots,p$ and \eqref{eq:contribution-jacobian-impact} gives
$\nabla_\xi\mathcal C_m(h,r)'$.
\end{proof}

\section*{Declaration of generative AI and AI-assisted technologies in the manuscript preparation process}

During the preparation of this work, the authors used ChatGPT and Claude
to assist with language editing and to improve the clarity of presentation.
After using these tools, the authors reviewed and edited the content as needed
and take full responsibility for the content of the published article.

\bibliographystyle{agsm}
\bibliography{references}

@article{dufour2006short,
  title={Short run and long run causality in time series: inference},
  author={Dufour, Jean-Marie and Pelletier, Denis and Renault, {\'E}ric},
  journal={Journal of Econometrics},
  volume={132},
  number={2},
  pages={337--362},
  year={2006},
  publisher={Elsevier}
}

@article{montiel2021local,
  title={Local projection inference is simpler and more robust than you think},
  author={Montiel Olea, Jos{\'e} Luis and Plagborg-M{\o}ller, Mikkel},
  journal={Econometrica},
  volume={89},
  number={4},
  pages={1789--1823},
  year={2021},
  publisher={Wiley Online Library}
}

@article{lutkepohl1990asymptotic,
  title={Asymptotic distributions of impulse response functions and forecast error variance decompositions of vector autoregressive models},
  author={L{\"u}tkepohl, Helmut},
  journal={The Review of Economics and Statistics},
volume={72},
  number={1},
  pages={116--125},
  year={1990},
  publisher={JSTOR}
}

@article{dufour1998short,
  title={Short run and long run causality in time series: theory},
  author={Dufour, Jean-Marie and Renault, Eric},
  journal={Econometrica},
  pages={1099--1125},
  volume={66},
  number={5},
  year={1998},
  publisher={JSTOR}
}

@article{wolf2021same,
  title={Local projections and {VAR}s estimate the same impulse responses},
  author={Plagborg-M{\o}ller, Mikkel and Wolf, Christian K},
  journal={Econometrica},
  volume={89},
  number={2},
  pages={955--980},
  year={2021},
  publisher={Wiley Online Library}
}

@article{box1975intervention,
  title={Intervention analysis with applications to economic and environmental problems},
  author={Box, George EP and Tiao, George C},
  journal={Journal of the American Statistical Association},
  volume={70},
  number={349},
  pages={70--79},
  year={1975},
  publisher={Taylor \& Francis}
}

@article{inoue2020uniform,
  title={The uniform validity of impulse response inference in autoregressions},
  author={Inoue, Atsushi and Kilian, Lutz},
  journal={Journal of Econometrics},
  volume={215},
  number={2},
  pages={450--472},
  year={2020},
  publisher={Elsevier}
}

@article{stock2018identification,
  title={Identification and estimation of dynamic causal effects in macroeconomics using external instruments},
  author={Stock, James H and Watson, Mark W},
  journal={Economic Journal},
  volume={128},
  number={610},
  pages={917--948},
  year={2018},
  publisher={Oxford University Press Oxford, UK}
}

@article{sims1980macroeconomics,
  title={Macroeconomics and reality},
  author={Sims, Christopher A},
  journal={Econometrica},
  volume={48},
  number={1},
  pages={1--48},
  year={1980},
  publisher={JSTOR}
}

@article{pesaran1998generalized,
  title={Generalized impulse response analysis in linear multivariate models},
  author={Pesaran, H Hashem and Shin, Yongcheol},
  journal={Economics letters},
  volume={58},
  number={1},
  pages={17--29},
  year={1998},
  publisher={Elsevier}
}

@article{kilian2011does,
  title={Does the Fed respond to oil price shocks?},
  author={Kilian, Lutz and Lewis, Logan T},
  journal={Economic Journal},
  volume={121},
  number={555},
  pages={1047--1072},
  year={2011},
  publisher={Oxford University Press Oxford, UK}
}

@article{bernanke1997systematic,
  title={Systematic monetary policy and the effects of oil price shocks},
  author={Bernanke, Ben S and Gertler, Mark and Watson, Mark},
  journal={Brookings Papers on Economic Activity},
  volume={1997},
  number={1},
  pages={91--157},
  year={1997},
  publisher={JSTOR}
}

@article{sims2006does,
  title={Does monetary policy generate recessions?},
  author={Sims, Christopher A and Zha, Tao},
  journal={Macroeconomic Dynamics},
  volume={10},
  number={2},
  pages={231--272},
  year={2006},
  publisher={Cambridge: Cambridge University Press}
}

@article{antolin2021structural,
  title={Structural scenario analysis with {SVAR}s},
  author={Antolin-Diaz, Juan and Petrella, Ivan and Rubio-Ramirez, Juan F},
  journal={Journal of Monetary Economics},
  volume={117},
  pages={798--815},
  year={2021},
  publisher={Elsevier}
}

@article{boudjellaba1992testing,
  title={Testing causality between two vectors in multivariate autoregressive moving average models},
  author={Boudjellaba, Hafida and Dufour, Jean-Marie and Roy, Roch},
  journal={Journal of the American Statistical Association},
  volume={87},
  number={420},
  pages={1082--1090},
  year={1992},
  publisher={Taylor \& Francis}
}

@article{dufour1993relationship,
  title={On the relationship between impulse response analysis, innovation accounting and {G}ranger causality},
  author={Dufour, Jean-Marie and Tessier, David},
  journal={Economics Letters},
  volume={42},
  number={4},
  pages={327--333},
  year={1993},
  publisher={Elsevier}
}

@article{rambachan2021common,
  title={When do common time series estimands have nonparametric causal meaning},
  author={Rambachan, Ashesh and Shephard, Neil},
  journal={Working Paper},
  year={2021}
}

@inproceedings{lucas1976econometric,
  title={Econometric policy evaluation: A critique},
  author={Lucas Jr, Robert E},
  booktitle={Carnegie-Rochester conference series on public policy},
  volume={1},
  pages={19--46},
  year={1976},
  organization={North-Holland}
}

@article{baron1986moderator,
  title={The moderator--mediator variable distinction in social psychological research: Conceptual, strategic, and statistical considerations.},
  author={Baron, Reuben M and Kenny, David A},
  journal={Journal of Personality and Social Psychology},
  volume={51},
  number={6},
  pages={1173},
  year={1986},
  publisher={American Psychological Association}
}

@article{mckay2023can,
  title={What Can Time-Series Regressions Tell Us About Policy Counterfactuals?},
  author={McKay, Alisdair and Wolf, Christian K},
  journal={Econometrica},
  volume={91},
  number={5},
  pages={1695--1725},
  year={2023},
  publisher={Wiley Online Library}
}

@article{bernanke1995inside,
  title={Inside the black box: the credit channel of monetary policy transmission},
  author={Bernanke, Ben S and Gertler, Mark},
  journal={Journal of Economic Perspectives},
  volume={9},
  number={4},
  pages={27--48},
  year={1995},
  publisher={American Economic Association}
}

@article{waggoner1999conditional,
  title={Conditional forecasts in dynamic multivariate models},
  author={Waggoner, Daniel F and Zha, Tao},
  journal={The Review of Economics and Statistics},
  volume={81},
  number={4},
  pages={639--651},
  year={1999},
  publisher={MIT Press 238 Main St., Suite 500, Cambridge, MA 02142-1046, USA journals~…}
}

@article{leeper2003modest,
  title={Modest policy interventions},
  author={Leeper, Eric M and Zha, Tao},
  journal={Journal of Monetary Economics},
  volume={50},
  number={8},
  pages={1673--1700},
  year={2003},
  publisher={Elsevier}
}

@article{hamilton2004comment,
  title={Oil shocks and aggregate macroeconomic behavior: the role of monetary policy},
  author={Hamilton, James D and Herrera, Ana Maria},
  journal={Journal of Money, Credit and Banking},
  pages={265--286},
volume={36},
  number={2},
  year={2004},
  publisher={JSTOR}
}

@article{blanchard1989dynamic,
  title={The Dynamic Effects of Aggregate Demand and Supply Disturbances},
  author={Blanchard, Olivier Jean and Quah, Danny},
  journal={American Economic Review},
  pages={655--673},
  year={1989},
  publisher={JSTOR}
}

@article{mertens2013dynamic,
  title={The dynamic effects of personal and corporate income tax changes in the United States},
  author={Mertens, Karel and Ravn, Morten O},
  journal={American Economic Review},
  volume={103},
  number={4},
  pages={1212--1247},
  year={2013},
  publisher={American Economic Association}
}

@article{kanzig2021macroeconomic,
  title={The macroeconomic effects of oil supply news: Evidence from OPEC announcements},
  author={K{\"a}nzig, Diego R},
  journal={American Economic Review},
  volume={111},
  number={4},
  pages={1092--1125},
  year={2021},
  publisher={American Economic Association 2014 Broadway, Suite 305, Nashville, TN 37203}
}

@article{imai2010identification,
  title={Identification, Inference and Sensitivity Analysis for Causal Mediation Effects},
  author={Imai, Kosuke and Keele, Luke and Yamamoto, Teppei},
  journal={Statistical Science},
  volume={25},
  number={1},
  pages={51--71},
  year={2010}
}

@book{vanderweele2015explanation,
  title={Explanation in causal inference: methods for mediation and interaction},
  author={VanderWeele, Tyler J},
  year={2015},
  publisher={Oxford University Press}
}

@book{mackinnon2012introduction,
  title={Introduction to statistical mediation analysis},
  author={MacKinnon, David},
  year={2012},
  publisher={Routledge}
}

@article{gilchrist2012credit,
  title={Credit spreads and business cycle fluctuations},
  author={Gilchrist, Simon and Zakraj{\v{s}}ek, Egon},
  journal={American Economic Review},
  volume={102},
  number={4},
  pages={1692--1720},
  year={2012},
  publisher={American Economic Association}
}

@article{uhlig2005effects,
  title={What are the effects of monetary policy on output? Results from an agnostic identification procedure},
  author={Uhlig, Harald},
  journal={Journal of Monetary Economics},
  volume={52},
  number={2},
  pages={381--419},
  year={2005},
  publisher={Elsevier}
}

@article{gali1999technology,
  title={Technology, employment, and the business cycle: do technology shocks explain aggregate fluctuations?},
  author={Gali, Jordi},
  journal={American Economic Review},
  volume={89},
  number={1},
  pages={249--271},
  year={1999},
  publisher={American Economic Association}
}

@article{romer2004new,
  title={A new measure of monetary shocks: Derivation and implications},
  author={Romer, Christina D and Romer, David H},
  journal={American Economic Review},
  volume={94},
  number={4},
  pages={1055--1084},
  year={2004},
  publisher={American Economic Association}
}

@article{jarocinski2020deconstructing,
  title={Deconstructing monetary policy surprises—the role of information shocks},
  author={Jaroci{\'n}ski, Marek and Karadi, Peter},
  journal={American Economic Journal: Macroeconomics},
  volume={12},
  number={2},
  pages={1--43},
  year={2020},
  publisher={American Economic Association 2014 Broadway, Suite 305, Nashville, TN 37203-2425}
}

@article{bauer2023risk,
  title={Risk appetite and the risk-taking channel of monetary policy},
  author={Bauer, Michael D and Bernanke, Ben S and Milstein, Eric},
  journal={Journal of Economic Perspectives},
  volume={37},
  number={1},
  pages={77--100},
  year={2023},
  publisher={American Economic Association 2014 Broadway, Suite 305, Nashville, TN 37203-2418}
}

@incollection{pearl2022direct,
  title={Direct and indirect effects},
  author={Pearl, Judea},
  booktitle={Probabilistic and causal inference: the works of Judea Pearl},
  pages={373--392},
  year={2022}
}

@incollection{kuersteiner2010granger,
  title={Granger-sims causality},
  author={Kuersteiner, Guido M},
  booktitle={Macroeconometrics and time series analysis},
  pages={119--134},
  year={2010},
  publisher={Springer}
}

@article{robins1992identifiability,
  title={Identifiability and exchangeability for direct and indirect effects},
  author={Robins, James M and Greenland, Sander},
  journal={Epidemiology},
  volume={3},
  number={2},
  pages={143--155},
  year={1992},
  publisher={LWW}
}

@article{robins2003semantics,
  title={Semantics of causal DAG models and the identification of direct and indirect effects},
  author={Robins, James M},
  journal={Highly structured stochastic systems},
  pages={70--82},
  year={2003},
  publisher={Oxford University PressOxford}
}

@article{chen2023direct,
  title={A direct approach to {Kilian--Lewis} style counterfactual analysis in vector autoregression models},
  author={Chen, Shiu-Sheng},
  journal={Journal of Applied Econometrics},
  volume={38},
  number={7},
  pages={1068--1076},
  year={2023},
  publisher={Wiley Online Library}
}

@article{robins1986new,
  title={A new approach to causal inference in mortality studies with a sustained exposure period—application to control of the healthy worker survivor effect},
  author={Robins, James},
  journal={Mathematical modelling},
  volume={7},
  number={9-12},
  pages={1393--1512},
  year={1986},
  publisher={Elsevier}
}

@inproceedings{avin2005identifiability,
  title={Identifiability of path-specific effects},
  author={Avin, Chen and Shpitser, Ilya and Pearl, Judea},
  booktitle={Proceedings of the 19th international joint conference on Artificial intelligence},
  pages={357--363},
  year={2005}
}

@article{dufour2025wald,
  title={Wald tests when restrictions are locally singular},
  author={Dufour, Jean-Marie and Renault, Eric and Zinde-Walsh, Victoria},
  journal={The Annals of Statistics},
  volume={53},
  number={2},
  pages={457--476},
  year={2025},
  publisher={Institute of Mathematical Statistics}
}

@article{acharya2016explaining,
  author  = {Acharya, Avidit and Blackwell, Matthew and Sen, Maya},
  title   = {Explaining Causal Findings Without Bias: Detecting and Assessing Direct Effects},
  journal = {American Political Science Review},
  year    = {2016},
  volume  = {110},
  number  = {3},
  pages   = {512--529}
}

@article{baumeister2018discretionary,
  author  = {Baumeister, Christiane and Kilian, Lutz and Zhou, Xiaoqing},
  title   = {Is the Discretionary Income Effect of Oil Price Shocks a Hoax?},
  journal = {Energy Journal},
  year    = {2018},
  volume  = {39},
  number  = {SI2},
  pages   = {117--137}
}

@article{kilian2008economic,
  author = {Kilian, Lutz}, title = {The Economic Effects of Energy Price Shocks},
  journal = {Journal of Economic Literature}, year = {2008}, volume = {46}, number = {4}, pages = {871--909}}

@article{hamilton1983oil,
  title={Oil and the macroeconomy since World War II},
  author={Hamilton, James D},
  journal={Journal of Political Economy},
  volume={91},
  number={2},
  pages={228--248},
  year={1983},
  publisher={The University of Chicago Press}
}

@misc{wegner2025transmission,
  author = {Wegner, Enrico and Lieb, Lenard and Smeekes, Stephan and Wilms, Ines},
  title  = {Transmission Channel Analysis in Dynamic Models},
  year   = {2025},
  note   = {arXiv:2405.18987v3}
}

@article{bernanke2004reply,
  author = {Bernanke, Ben S. and Gertler, Mark and Watson, Mark W.},
  title = {Reply to ``Oil Shocks and Aggregate Economic Behavior: The Role of Monetary Policy''},
  journal = {Journal of Money, Credit and Banking}, year = {2004}, volume = {36}, number = {2}, pages = {287--291}}

@article{leduc2004quantitative,
  author = {Leduc, Sylvain and Sill, Keith},
  title = {A Quantitative Analysis of Oil-Price Shocks, Systematic Monetary Policy, and Economic Downturns},
  journal = {Journal of Monetary Economics}, year = {2004}, volume = {51}, number = {4}, pages = {781--808}}

@article{carlstrom2006oil,
  author = {Carlstrom, Charles T. and Fuerst, Timothy S.},
  title = {Oil Prices, Monetary Policy, and Counterfactual Experiments},
  journal = {Journal of Money, Credit and Banking}, year = {2006}, volume = {38}, number = {7}, pages = {1945--1958}}

@article{jentsch2019asymptotically,
  author  = {Jentsch, Carsten and Lunsford, Kurt G.},
  title   = {The Dynamic Effects of Personal and Corporate Income Tax Changes in the United States: Comment},
  journal = {American Economic Review},
  year    = {2019},
  volume  = {109},
  number  = {7},
  pages   = {2655--2678},
  doi     = {10.1257/aer.20162011}
}

@article{edelstein2009sensitive,
  author  = {Edelstein, Paul and Kilian, Lutz},
  title   = {How Sensitive Are Consumer Expenditures to Retail Energy Prices?},
  journal = {Journal of Monetary Economics},
  year    = {2009},
  volume  = {56},
  number  = {6},
  pages   = {766--779},
  doi     = {10.1016/j.jmoneco.2009.06.001}
}

@book{imbens2015causal,
  title={Causal Inference for Statistics, Social, and Biomedical Sciences: An Introduction},
  author={Imbens, Guido W and Rubin, Donald B},
  year={2015},
  publisher={Cambridge University Press}
}

@article{vanderweele2011controlled,
  title={Controlled Direct and Mediated Effects: Definition, Identification and Bounds},
  author={VanderWeele, Tyler J},
  journal={Scandinavian Journal of Statistics},
  volume={38},
  number={3},
  pages={551--563},
  year={2011},
  doi={10.1111/j.1467-9469.2010.00722.x}
}

@article{vansteelandt2017interventional,
  title={Interventional Effects for Mediation Analysis with Multiple Mediators},
  author={Vansteelandt, Stijn and Daniel, Rhian M},
  journal={Epidemiology},
  volume={28},
  number={2},
  pages={258--265},
  year={2017},
  doi={10.1097/EDE.0000000000000596}
}

\end{document}